\PassOptionsToPackage{pdfpagelabels=false}{hyperref}
\RequirePackage{fix-cm}
\documentclass[fleqn,usenatbib]{mnras}

\usepackage{mathptmx}

\usepackage[T1]{fontenc}
\usepackage{ae,aecompl}

\usepackage{graphicx}	
\usepackage{amsmath}	
\usepackage{amssymb}	
\usepackage{epstopdf} 
\usepackage[normal]{threeparttable} 
\usepackage{pstricks}
\usepackage{float}
\usepackage{subcaption}
\usepackage{longtable}

\title[The 2017 May 20$^{th}$ stellar occultation by 2002 GZ$_{32}$]{The 2017 May 20$^{\rm th}$ stellar occultation by the elongated centaur (95626) 2002 GZ$_{32}$}

\author[P. Santos-Sanz et al.]
{\Large P. Santos-Sanz,$^{1}$\thanks{E-mail: psantos@iaa.es (IAA-CSIC)}
J. L. Ortiz,$^{1}$
B. Sicardy,$^{2}$
G. Benedetti-Rossi,$^{2,3,4}$
N. Morales,$^{1}$ 
E. Fern\'{a}ndez-Valenzuela,$^{5}$
\newauthor
\Large
R. Duffard,$^{1}$
R. Iglesias-Marzoa,$^{6,7}$
J.L. Lamadrid,$^{6}$
N. Ma\'{i}cas,$^{6}$
L. P\'{e}rez,$^{8}$ 
K. Gazeas,$^{9}$  
J.C. Guirado,$^{10,11}$ 
\newauthor
\Large
V. Peris,$^{10}$ 
F.J. Ballesteros,$^{10}$ 
F. Organero,$^{12}$  
L. Ana-Hern\'{a}ndez,$^{12}$  
F. Fonseca,$^{12}$ 
A. Alvarez-Candal,$^{3}$
\newauthor
\Large
Y. Jim\'{e}nez-Teja,$^{3}$
M. Vara-Lubiano,$^{1}$	
F. Braga-Ribas,$^{13,2,3,4}$
J.I.B. Camargo,$^{3,4}$
J. Desmars,$^{14,15}$
\newauthor
\Large
M. Assafin,$^{4,16}$
R. Vieira-Martins,$^{3,4}$
J. Alikakos,$^{17}$
M. Boutet,$^{18}$
M. Bretton,$^{19}$
A. Carbognani,$^{20}$
\newauthor
\Large
V. Charmandaris,$^{21,22}$
F. Ciabattari,$^{23}$
P. Delincak,$^{24}$
A. Fuambuena Leiva,$^{25}$
H. Gonz\'{a}lez,$^{26}$
T. Haymes,$^{27}$
\newauthor
\Large
S. Hellmich,$^{28}$
J. Horbowicz,$^{29}$
M. Jennings,$^{30}$
B. Kattentidt,$^{31}$
Cs. Kiss,$^{32,33}$
R. Kom\v{z}\'{i}k,$^{34}$
J. Lecacheux,$^{2}$
\newauthor
\Large
A. Marciniak,$^{29}$
S. Moindrot,$^{35}$
S. Mottola,$^{28}$
A. Pal,$^{32}$
N. Paschalis,$^{36}$
S. Pastor,$^{37}$
C. Perello,$^{38,39}$
\newauthor
\Large
T. Pribulla,$^{34,40,41}$
C. Ratinaud,$^{42}$
J.A. Reyes,$^{37}$
J. Sanchez,$^{18}$
C. Schnabel,$^{38,39}$
A. Selva,$^{38,39}$
F. Signoret,$^{25}$
\newauthor
\Large
E. Sonbas,$^{43}$
V. Al\'{i}-Lagoa$^{44}$\\
$^{1}$Instituto de Astrof\'{i}sica de Andaluc\'{i}a, IAA-CSIC, Glorieta de la Astronom\'{i}a s/n, 18008 Granada, Spain\\
$^{2}$LESIA, Observatoire de Paris, PSL Research University, CNRS, Sorbonne Universit\'e, Univ. Paris Diderot, Sorbonne Paris Cit\'e, France\\
$^{3}$Observat\'{o}rio Nacional/MCTIC, Rio de Janeiro, Brazil\\
$^{4}$Laborat\'{o}rio Interinstitucional de e-Astronomia (LIneA) and INCT do e-Universo, Brazil\\
$^{5}$Florida Space Institute, University of Central Florida, Orlando, USA\\
$^{6}$Centro de Estudios de F\'{i}sica del Cosmos de Arag\'{o}n, Teruel, Spain\\
$^{7}$Departamento de Astrof\'{i}sica, Universidad de La Laguna, La Laguna, Spain\\
$^{8}$Observatorio de Allariz, Spain\\
$^{9}$Section of Astrophysics, Astronomy and Mechanics, Department of Physics, National and Kapodistrian University of Athens, Athens, Greece\\
$^{10}$Observatori Astron\`{o}mic de la Universitat de Val\`{e}ncia, Paterna, Spain\\
$^{11}$Departament d'Astronomia i Astrof\'{i}sica, Universitat de Val\`{e}ncia, Burjassot, Spain\\
$^{12}$Observatorio Astron\'{o}mico La Hita, Toledo, Spain\\
$^{13}$Federal University of Technology--Paran\'{a} (UTFPR/Curitiba), Brazil\\
$^{14}$Institut Polytechnique des Sciences Avanc\'ees IPSA, Ivry-sur-Seine, France\\
$^{15}$IMCCE, Observatoire de Paris, PSL Research University, CNRS, Sorbonne Universit\'es, UPMC Univ Paris 06, Univ. Lille, Paris, France\\
$^{16}$Observat\'{o}rio do Valongo/UFRJ, Brazil\\
$^{17}$Institute for Astronomy, Astrophysics, Space Applications and Remote Sensing, National Observatory of Athens, Penteli, Greece\\
$^{18}$Latrape Observatory, Toulouse, France\\
$^{19}$Observatoire des Baronnies Proven\c{c}ales, Moydans, France\\
$^{20}$INAF-Osservatorio di Astrofisica e Scienza dello Spazio (OAS), Bologna, Italy\\
$^{21}$Department of Physics, University of Crete, Heraklion, Greece\\
$^{22}$Institute of Astrophysics, FORTH, GR-71110, Heraklion,  Greece\\
$^{23}$Osservatorio Astronomico di Monte Agliale, Via Cune Motrone, Borgo a Mozzano, Italy\\
$^{24}$PDlink Observatory, Cadca, Slovakia\\
$^{25}$T\'{e}lescope L\'{e}onard de Vinci, Antibes, France\\
$^{26}$Observatorio Astron\'{o}mico de Forcarei (OAF), Spain\\
$^{27}$Smithy Observatory, Hill Rise, Knowl Hill Common, UK, for the British Astronomical Association\\
$^{28}$German Aerospace Center (DLR), Institute of Planetary Research, Berlin, Germany\\
$^{29}$Astronomical Observatory Institute, Faculty of Physics, A. Mickiewicz University, Sloneczna 36, 60-286 Pozna\'{n}, Poland\\
$^{30}$Hamsey Green Observatory, London, UK\\
$^{31}$Neutraubling, Regensburg, Germany\\
$^{32}$Konkoly Observatory, Research Centre for Astronomy and Earth Sciences, Konkoly Thege 15-17, H-1121 Budapest, Hungary\\
$^{33}$ELTE E\"otv\"os Lor\'and University, Institute of Physics, Budapest, Hungary\\
$^{34}$Astronomical Institute, Slovak Academy of Sciences, Tatransk\'{a} Lomnica, Slovakia\\
$^{35}$Observatoire de Puimichel, France\\
$^{36}$Nunki Observatory, Skiathos Island, Greece\\
$^{37}$La Murta Observatory--Astromurcia, Murcia, Spain\\
$^{38}$Agrupaci\'{o} Astron\`{o}mica de Sabadell, Barcelona, Spain\\
$^{39}$International Occultation Timing Association--European Section (IOTA-ES), Germany\\
$^{40}$ELTE Gothard Astrophysical Observatory, 9700 Szombathely, Szent Imre h.u. 112, Hungary\\
$^{41}$MTA-ELTE Exoplanet Research Group, 9700 Szombathely, Szent Imre h.u. 112, Hungary\\
$^{42}$Landehen Observatory, France\\
$^{43}$University of Adiyaman, Department of Physics, 02040 Adiyaman, Turkey\\
$^{44}$Max Planck Institut f\"ur extraterrestrische Physik (MPE), Garching, Germany}

\date{Accepted XXX. Received YYY; in original form ZZZ}

\pubyear{2019}

\begin{document}
\label{firstpage}
\pagerange{\pageref{firstpage}--\pageref{lastpage}}
\maketitle

\begin{abstract}

We predicted a stellar occultation of the bright star Gaia DR1 4332852996360346368 (UCAC4 385-75921) (m$_{\rm V}$= 14.0 mag) by the centaur 2002 GZ$_{32}$ for 2017 May 20$^{\rm th}$. Our latest shadow path prediction was favourable to a large region in Europe. Observations were arranged in a broad region inside the nominal shadow path. Series of images were obtained with 29 telescopes throughout Europe and from six of them (five in Spain and one in Greece) we detected the occultation. This is the fourth centaur, besides Chariklo, Chiron and Bienor, for which a multi-chord stellar occultation is reported. By means of an elliptical fit to the occultation chords we obtained the limb of 2002 GZ$_{32}$ during the occultation, resulting in an ellipse with axes of 305 $\pm$ 17 km $\times$ 146 $\pm$ 8 km. From this limb, thanks to a rotational light curve obtained shortly after the occultation, we derived the geometric albedo of 2002 GZ$_{32}$ ($p_{\rm V}$ = 0.043 $\pm$ 0.007) and a 3-D ellipsoidal shape with axes 366 km $\times$ 306 km $\times$ 120 km. This shape is not fully consistent with a homogeneous body in hydrostatic equilibrium for the known rotation period of 2002 GZ$_{32}$. The size (albedo) obtained from the occultation is respectively smaller (greater) than that derived from the radiometric technique but compatible within error bars. No rings or debris around 2002 GZ$_{32}$ were detected from the occultation, but narrow and thin rings cannot be discarded.

\end{abstract}



\section{Introduction}

Although no official definition exists, centaurs such as (95626) 2002 GZ$_{32}$, are Solar System small bodies that orbit the Sun between the orbits of Jupiter and Neptune. Due to their large distance to the Sun, centaurs, together with trans-neptunian objects (TNOs) are thought to be the least evolved and the most pristine bodies in the Solar System, at least from the composition point of view. This means that they retained relevant information on the composition materials and physical conditions of the primitive solar nebula. Therefore, the study of these bodies reveals plenty of information on the origin and evolution of the Solar System since its initial phases. Additionally, centaurs are considered the progenitors of the short period comets, mainly the Jupiter family comets, and it is believed that they escaped from the TNO population referred to as the Scattered Disc Objects (SDOs) \citep[e.g.,][]{Fernandez2018}.

The first centaur discovered was Chiron, found by Charles Kowal in 1977 \citep{kowal89}. Despite the fact that more than 40 years have elapsed since the first member of this population was discovered, our knowledge about the physical properties of the centaur population is still scarce, mainly due to the large distance, low geometric albedo and faintness of these bodies. Centaurs are thought to be mostly composed of a mixture of ices and rocks and most of the members of the population probably share similar composition to that of comets \citep[e.g.,][]{Barucci2011}.

The centaur (95626) 2002 GZ$_{32}$ was discovered on 2002 April 13$^{\rm th}$ from Mauna Kea Observatory. It presents a rotational period of 5.80 $\pm$ 0.03 h with a peak-to-peak amplitude of 0.15 magnitudes according to \cite{Dotto2008}. A variable presence of water ice on the surface of this centaur has been reported \citep[][see Table \ref{OrbPhys_GZ32}]{Barkume2008}, and the thermal emission from its surface has been detected with Spitzer/MIPS in July 2007 and later with Herschel/PACS in August 2010. Using the Spitzer and Herschel thermal measurements and $H_{\rm V}$ = 7.37 $\pm$ 0.09 mag \cite{Duffard2014} obtained the area-equivalent diameter ($D_{\rm eq} = 237 \pm 8$ km) and the geometric albedo at V-band ($p_{\rm V} = 0.037 \pm 0.004$) of 2002 GZ$_{32}$ via the radiometric technique. \cite{Lellouch2017} updated these radiometric results adding ALMA observations to the Herschel and Spitzer data, obtaining an area-equivalent diameter of 237$^{+12}_{-11}$ km and a geometric albedo of 0.036$^{+0.006}_{-0.005}$. The large radiometric size derived means that 2002 GZ$_{32}$ is one of the largest objects within the centaur population, similar in size to Chariklo and Chiron. A summary of the orbital elements and most relevant physical characteristics of 2002 GZ$_{32}$ is shown in Table \ref{OrbPhys_GZ32}, including an updated absolute magnitude ($H_{\rm V}$ = 7.39 $\pm$ 0.06 mag) recently obtained by \cite{Alvarez-Candal2019}. This new $H_{\rm V}$ is used for the calculations in this work.

\begin{table*}
	\centering
	\caption{Orbital elements and most relevant physical characteristics of the centaur 2003 GZ$_{32}$.}
	\label{OrbPhys_GZ32}
	\resizebox{\textwidth}{!}{%
	\begin{tabular}{ccccccccccc}   
		\hline	    
\textbf{a} & \textbf{q} & \textbf{e} & \textbf{i} & \textbf{$H_{\rm V}$}$^{a}$ & \textbf{P}$^{b}$ & \textbf{$\Delta \rm m$}$^{c}$ & \textbf{D} & \textbf{$p_{\rm V}$} & \textbf{Spec. Slope}$^{f}$ & \textbf{Ices}$^{f}$ \\

\textbf{[AU]} & \textbf{[AU]} &  & \textbf{[deg]} & \textbf{[mag]} & \textbf{[h]} & \textbf{[mag]} & \textbf{[km]} &  & \textbf{\%/1000 \AA} &  \\
\hline
23.02 & 18.01 & 0.218 & 15.035 & 7.39 $\pm$ 0.06  & 5.80 $\pm$ 0.03 & 0.13 $\pm$ 0.01 & 237 $\pm$ 8 $^{d}$ 	& 0.037 $\pm$ 0.004 $^{d}$ 		 & 16.9 $\pm$ 0.1 & Water? \\
      &       &       &  	   &   					&  	   &  				 & 237$^{+12}_{-11}$ $^{e}$ & 0.036$^{+0.006}_{-0.005}$ $^{e}$ &      & 		\\
\hline
	\end{tabular}}
\begin{tablenotes}
      \small
   	  \item \textbf{Orbital elements} --a, q, e and i-- from JPL-Horizons. \textbf{$H_{\rm V}$} is the absolute magnitude at V-band. \textbf{P} is the rotational period. \textbf{$\Delta \rm m$} is the peak-to-peak amplitude obtained from the rotational light curve. \textbf{D} and \textbf{$p_{\rm V}$} are, respectively, the area-equivalent diameter and the geometric albedo at V-band, both obtained from radiometric techniques. \textbf{Spec. Slope} is the visible spectral slope. \textbf{Ices} indicates the possible icy species detected from spectra. All uncertainties are 1$\sigma$. 
      \item \textbf{References:} $^{a}$\cite{Alvarez-Candal2019}, $^{b}$\cite{Dotto2008}, $^{c}$This work, $^{d}$\cite{Duffard2014},  $^{e}$\cite{Lellouch2017}, $^{f}$\cite{Barkume2008}.
\end{tablenotes}
\end{table*}

Stellar occultations by only five centaurs --Asbolus, Bienor, Chariklo, Chiron and Echeclus-- have been recorded so far \citep{Ortiz2020}, and only for three of them (Chariklo, Chiron and, very recently, Bienor --Morales et al., private communication--) multi-chord{\footnote {We define `multi-chord occultation' as an occultation detected from more than one site (i.e. at least two positive detections).}} occultations were obtained. A ring system around the centaur Chariklo \citep{Braga-Ribas2014} and a possible similar structure --still under discussion-- around the centaur Chiron \citep{Ortiz2015,Ruprecht2015,Sickafoose2020} have been reported from these multi-chord occultations. This highlights the relevance of observing stellar occultations by the largest centaurs: apart from determining their sizes, shapes, albedos and other physical properties with high accuracy, detecting new rings might be possible. The discovery of rings around small bodies has opened a new avenue of research within the planetary science. The finding that the TNO and dwarf planet Haumea also has a ring \citep{Ortiz2017} has spurred interest on this topic even more. The possible relationship and physical properties behind the rings around centaurs and TNOs is also a new and exciting way of research that has just started to be explored \citep[e.g.,][]{Sicardy2019a,Sicardy2019b}. The large size of 2002 GZ$_{32}$ ($\sim$ 240 km), the likely presence of water ice on its surface (see Table \ref{OrbPhys_GZ32}) and its short rotation period --all these properties also present in Chariklo and Chiron-- make this centaur a potential candidate to have a ring system.

In this paper we present the first determination of the size, albedo, projected shape and search for rings around 2002 GZ$_{32}$ based on the multichord stellar occultation of 2017 May 20$^{\rm th}$. We also present an updated rotational light curve for 2002 GZ$_{32}$ obtained with the 1.23-m telescope at Calar Alto Observatory (CAHA) in Spain close to the occultation date. This rotational light curve is relevant to analyze the stellar occultation results in terms of a three dimensional shape. The paper is organized as follows: in Section \ref{obs} we describe the astrometric observations performed to predict the stellar occultation, the stellar occultation observations themselves, and the observations to determine the rotational light curve; in Section \ref{analysis} we detail the reduction and analysis of the stellar occultation data; the limb fit, diameter, 3D shape, albedo, density and search for rings are presented and discussed in Section \ref{results}; finally, we present our conclusions in Section \ref{conclu}.



\section{Observations}
\label{obs}

\subsection{Predictions}
\label{pred}

Due to the growing interest in recording stellar occultations by TNOs and centaurs, we have been performing intensive astrometric and photometric campaigns since 2010 in order to predict and to observe these occultations with the aim to obtain relevant physical information (e.g., size, shape, albedo, density, rings) of these bodies --see \cite{Ortiz2020} for a review--. In the course of our astrometric campaigns we found out that the centaur 2002 GZ$_{32}$ would occult a V= 14.0 mag star on 2017 May 20$^{\rm th}$. The original prediction of the occultation was obtained around one year prior to the event using our own astrometric observations and the offsets with respect to the orbit JPL\#20 as described below. To obtain this prediction we took 14 images in 2 $\times$ 2 binning mode of 2002 GZ$_{32}$ on May 30 and June 14, 23 and 26, 2016 with the 4k $\times$ 4k IO:O camera of the Liverpool 2-m telescope in Roque de los Muchachos Observatory in La Palma (Spain). The detailed setup of these observations, including weather condictions and other related information, are shown in Table \ref{ObsSummary}. Bias and sky flat-field frames were taken each night to calibrate the images.

We also took 18 images of the centaur in 2 $\times$ 2 binning mode on May 16, 2017 with the 4k $\times$ 4k DLR-MKIII CCD of the 1.23-m Calar Alto Observatory telescope in Almer\'ia (Spain). More details of this observing run are included in Table \ref{ObsSummary}. Bias frames and twilight sky flat-field frames were taken to calibrate the images. The astrometry obtained from these Calar Alto images allowed us to obtain an independent `last minute' prediction using the offsets with respect to the orbit JPL\#23.

\begin{table*}
	\centering
	\caption{Summary of the astrometric observing campaigns}
	\label{ObsSummary}
	\resizebox{\textwidth}{!}{%
	\begin{tabular}{ccccccccccccccc}   
		\hline	    
\textbf{Telescope} & \textbf{Date} & \textbf{CCD} & \textbf{Scale} & \textbf{FOV} & \textbf{Filter} & \textbf{Exp.} & \textbf{N} & \textbf{Seeing} & \textbf{SNR} & \textbf{1$\sigma$(RA)} & \textbf{1$\sigma$(Dec)} & \textbf{Offset(RA)} & \textbf{Offset(Dec)} & \textbf{JPL\#}\\
\hline
\textbf{Liverpool 2-m} & May-Jun 2016 & 4k $\times$ 4k & 0.15$''$/pix & 10$'$ $\times$ 10$'$  & R-Sloan & 400 s & 14 & 1.3$''$-2$''$ & 50-100 & 18 mas & 30 mas & -7 mas & +19 mas & 20 \\
\textbf{Calar Alto 1.23-m} & May 2017 & 4k $\times$ 4k & 0.32$''$/pix & 21$'$ $\times$ 21$'$ & R-Johnson & 400 s & 18 & $\sim$ 2$''$ & $\sim$ 45 & 35 mas & 34 mas & -74 mas & +30 mas & 23 \\

\hline
	\end{tabular}}
\begin{tablenotes}
      \small
   	\item \textbf{Telescope} indicates the telescope used during the observing run. \textbf{Date} are the observing dates (more details in the text of Section \ref{pred}). \textbf{CCD} indicates the size of the detector. \textbf{Scale} is the unbinned image scale of the instrument in arcseconds per pixel. \textbf{FOV} is the field of view of the instrument in arcminutes $\times$ arcminutes. \textbf{Filter} is the filter used. \textbf{Exp.} is the exposure time in seconds. \textbf{N} is the number of images. \textbf{Seeing} is the seeing variation or average seeing during the observing runs. \textbf{SNR} is the signal to noise ratio of the object. \textbf{1$\sigma$(RA)} is the average 1$\sigma$ uncertainty in right ascension of the astrometry in milliarcseconds. \textbf{1$\sigma$(Dec)} is the average 1$\sigma$ uncertainty in declination of the astrometry in milliarcseconds. \textbf{Offset(RA)} is the offset in right ascension with respect to the corresponding JPL orbit expressed in milliarcseconds. \textbf{Offset(Dec)} is the offset in declination with respect to the corresponding JPL orbit expressed in milliarcseconds. \textbf{JPL\#} is the JPL orbit used to calculate the offsets.
 
\end{tablenotes}
\end{table*}

\begin{figure}
    \includegraphics[width=\columnwidth]{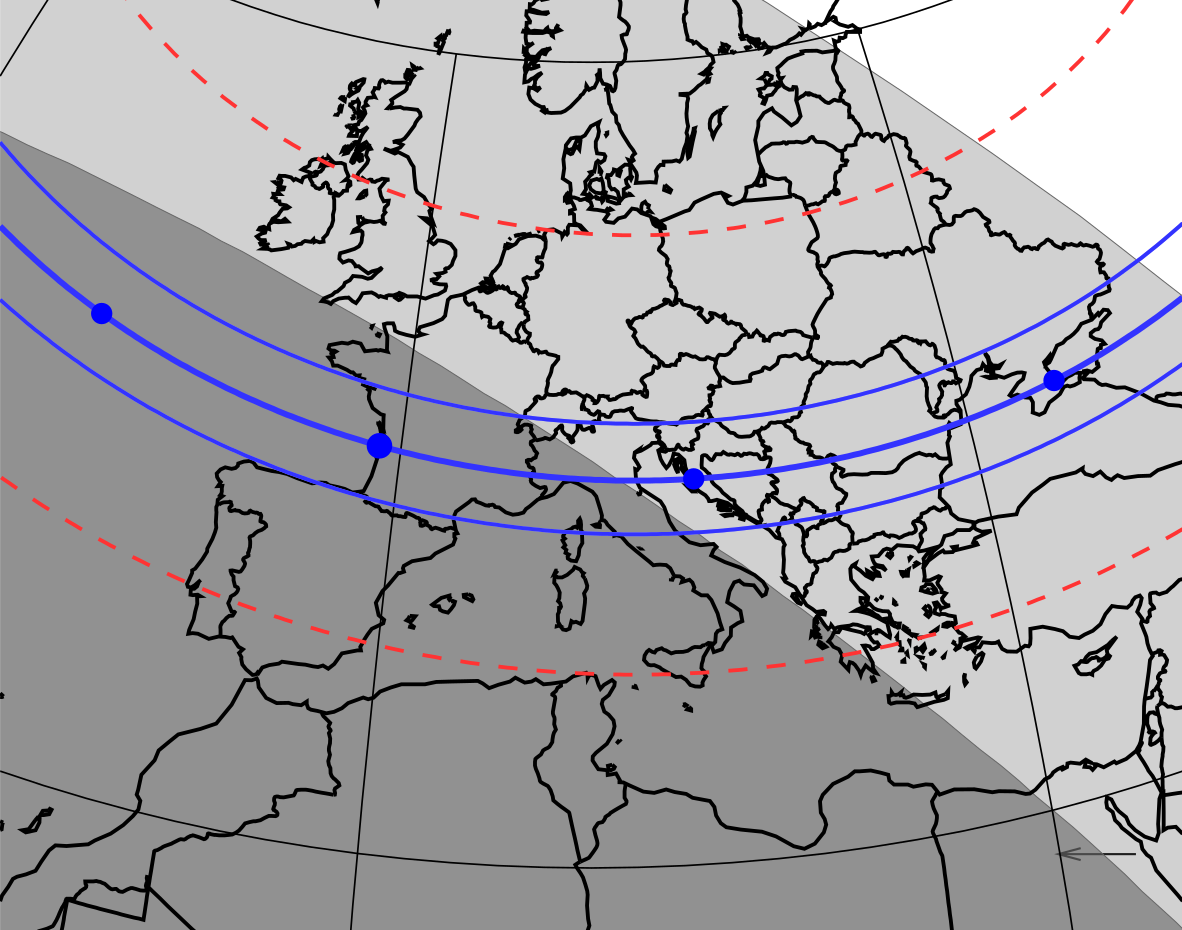}
	\includegraphics[width=\columnwidth]{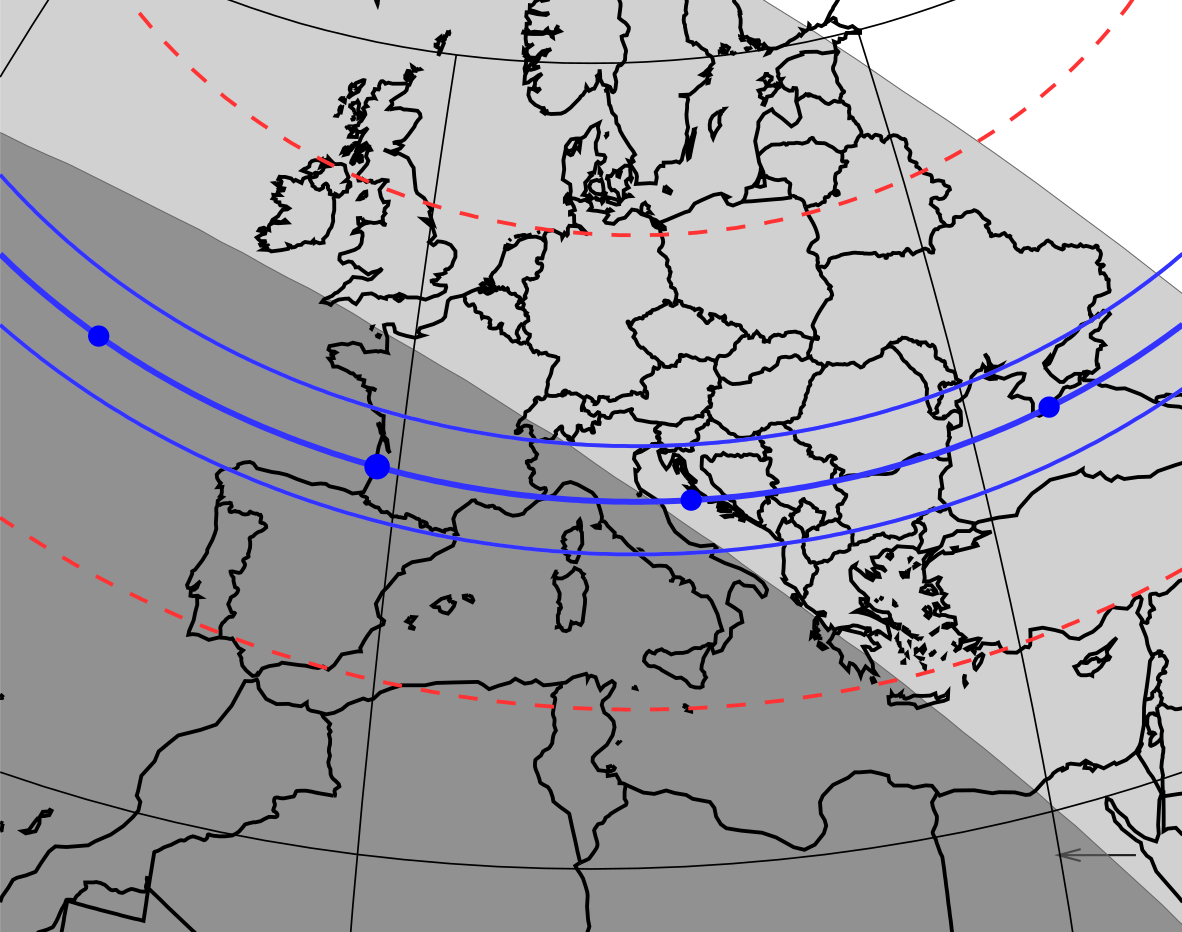}
	
    \caption{Two independent predictions of the occultation using Gaia DR1 star catalogue \citep{Gaia2016} and relative astrometry of 2002 GZ$_{32}$ with respect to the occulted star obtained from the 1.23-m telescope in Calar Alto Observatory, Almeria, Spain (top panel), and from the 2-m Liverpool Telescope in Canary Islands, Spain (bottom panel). The thick blue line indicates the centerline of the shadow path and the thin blue lines the limits of the shadow. Blue dots represent the position of the center of the body spaced every 1 minute. The 1-$\sigma$ precision along the path is represented by the red dotted line. This corresponds to 840 km in the top panel and 740 km in the bottom panel. Both values are projected distances on Earth's surface. The 1-$\sigma$ precision in time of the prediction is of 20 s for the top panel and of 10 s for the bottom panel. The arrow in the right bottom shows the direction of the shadow motion. The width of the shadow path in the predictions is assumed to be the diameter of 2002 GZ$_{32}$ derived from Herschel and Spitzer thermal data \citep[D = 237 km, according to][]{Duffard2014}. The projected distance on Earth surface between the central lines of the two predictions is $\sim$ 100 km. The real shadow path of the occultation was $\sim$ 580 km south of the Calar Alto prediction and $\sim$ 480 km south of the Liverpool prediction (see Figure \ref{shadowpath}) but within the estimated uncertainties.}
    \label{predictmap}
\end{figure}

The images acquired during the aforementioned observing campaigns were astrometrically solved using Gaia DR1 \citep{Gaia2016} which was the most precise astrometric catalogue available at the time of this occultation, Gaia DR2 \citep{Gaia2018} was not yet available. Proper motions of the occulted star were accounted for using the HSOY catalogue \citep{Altmann2017}. The predictions from the two telescopes were obtained using the coordinates of the occulted star (Gaia DR1 4332852996360346368), derived as described above, and the relative astrometry (offsets) of 2002 GZ$_{32}$ with respect to the Jet Propulsion Laboratory orbits JPL\#20 (for the 2016 data from the Liverpool Telescope) and JPL\#23 (for the 2017 data from Calar Alto). The average 1$\sigma$ uncertainties in the astrometry obtained from both telescopes and the offsets with respect to the orbits JPL\#20 and JPL\#23 are shown in Table \ref{ObsSummary}. These uncertainties translate to 1$\sigma$ uncertainties of 740 km projected on Earth surface (cross-track) and 10 s in time (along-track) for the Liverpool data and of 840 km projected on Earth surface and 20 s in time for the Calar Alto data. The prediction map obtained from the Liverpool telescope astrometry is shown in the bottom panel of Figure \ref{predictmap} and the map obtained from the CAHA 1.23-m telescope astrometry is shown in the top panel of the same Figure. It is important to note that the two predictions are independent of each other because they are based on independent offsets obtained from the two datasets. Nevertheless, both predictions were compatible within error bars and they have a difference between their central lines of $\sim$ 100 km in projected distance on Earth surface.

As the original prediction from the Liverpool telescope and the `last minute' one from the Calar Alto telescope were consistent and compatible with each other, we decided to alert our team members and a large list of professional and amateur European collaborators. Six positive detections, at five well-separated sites, and two near misses were finally obtained (see Section \ref{OccObs}).

\subsection{Stellar Occultation}
\label{OccObs}

\begin{figure}
	\includegraphics[width=\columnwidth]{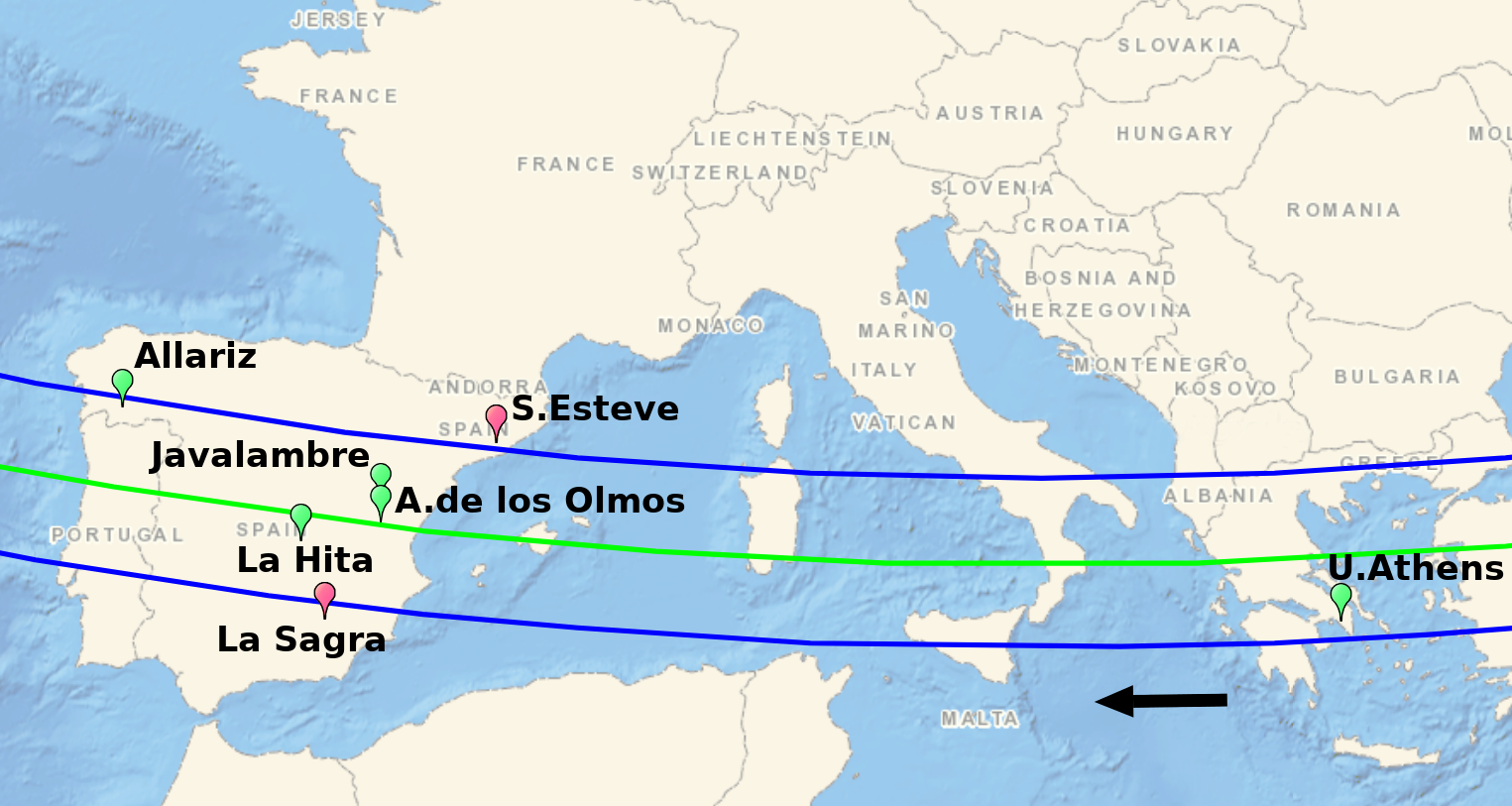}
    \caption{Map of the real shadow path (i.e. the reconstructed ellipse fit) of the stellar occultation by 2002 GZ$_{32}$ including the sites (green marks) from where the occultation was recorded (Allariz, Javalambre, Aras de los Olmos and La Hita in Spain, and University of Athens in Greece) and the negative detections (red marks) closest to the shadow path (Sant Esteve and La Sagra in Spain). The green line indicates the centerline of the shadow path and the blue lines the limits of the shadow. Direction of the shadow is shown by the black arrow. Map credit: \url{https://www.gpsvisualizer.com/} and Australian Topography (\textsuperscript{\textcopyright}Commonwealth of Australia --Geoscience Australia-- 2016. Creative Commons Attribution 4.0 International Licence).}
    \label{shadowpath}
\end{figure}

Series of FITS images or video observations were obtained with 29 telescopes on May 20$^{\rm th}$ 2017, and from six of them (five in Spain and one in Greece) we recorded the disappearance and reappearance of the star. We obtained light curves from those six positive observations at five sites (with two telescopes located at La Hita Observatory) that showed deep drops of different duration around the predicted occultation time. Light curves from the other observing sites were also obtained, putting special care in the negative detections at the two locations closest to the shadow path (Sant Esteve and La Sagra Observatory, both in Spain), because they can notably constrain the projected shape of 2002 GZ$_{32}$ (see Figure \ref{shadowpath} and Table \ref{ObservDetails}). Apart from these six positive detections (from five sites) and two closest non-detections, the occultation was negative from 21 other telescopes located in Spain (5), UK (2), France (6), Italy (2), Germany (1), Poland (1), Hungary (1), Slovakia (2) and Turkey (1). Two other observatories in Greece, both located under the shadow path, missed the occultation due to bad weather conditions and technical problems, respectively. Table \ref{SummaryObserv} lists all telescopes that participated in the occultation campaign. Robust and reliable clock synchronizations were used in all the observing sites (i.e., Internet Network Time Protocol servers --NTP-- or GPS-based Video Time inserters --VTI--), the acquisition times of each image were inserted on the corresponding image header. All CCD images were obtained from around fifteen minutes before and after the predicted occultation time. No filters were used in order to maximize the SNR of the photometric data.

The light curves of the occultation (flux versus time) were obtained using our own daophot-based routines \citep{Stetson1987}, coded in the Interactive Data Language (IDL), after the usual bias and sky flat-field calibration of the individual images (i.e., subtracting a median bias and dividing by a median flat-field). Relative aperture photometry of the occulted star (which, around the occultation time, is the combination of the star flux and the 2002 GZ$_{32}$ flux) was then obtained on the images using the stars present in the FOV as comparison stars in order to minimize variations due to different seeing conditions and atmospheric transparency fluctuations. The chosen aperture radii and comparison stars minimized the flux dispersion of the occulted star before and after the occultation. Finally, the combined flux of the occulted star and 2002 GZ$_{32}$ versus time was obtained for each observing site. From the five locations cited above the obtained light curves show deep drops in flux at the predicted occultation times (see Figure \ref{OccLCs}). Detailed information about the observatories (telescopes, detectors, exposure times, etc) from where the positive detections were obtained are shown in Table \ref{ObservDetails}, together with the stations closest to the object that reported negative detection.

Note that the observations obtained in video format were converted to FITS images prior to their analysis. The only positive observation acquired in video was the La Hita Observatory T-40cm one, which was used to confirm the positive light curve from the same site using the 77-cm telescope and a CCD detector (see Table \ref{ObservDetails}).  

Further details on the analysis of the stellar occultation light curves are given in Section \ref{analysis}.

\begin{table*}
	\centering
	\caption{Observation details of the stellar occultation by 2002 GZ$_{32}$ including positive and negative detections used to constrain the shape of the object. $\sigma_{\rm flux}$ is the flux dispersion, i.e. the standard deviation of the normalized flux (outside of the occultation in the case of positive detections). All the positive detections used to obtain the limb fit were based on the computer's system time and the Network Time Protocol (NTP) time synchronized few minutes before the start of the observation.}
	\label{ObservDetails}
	\resizebox{\textwidth}{!}{%
	\begin{tabular}{cccccc} 
		\hline
\textbf{Observatory name} 	& \textbf{Longitude (E)} & \textbf{Telescope characteristics}  & \textbf{Exposure time} & \textbf{Observer(s)} & \textbf{Detection}\\
\textbf{(City, Country)	}	& \textbf{Latitude (N)}  & \textbf{Detector/Instrument} & \textbf{Cycle time}    & 					 & \textbf{$\sigma_{\rm flux}$} \\
\textbf{IAU code} 		 	& \textbf{Altitude (m)}  & 					   & \textbf{(seconds)} 	   & &\\
 \hline
 \hline
Allariz Observatory & -07$^\circ$ 46' 13.0''	  & D= 0.25 m, f= 1200 mm 	& 1.5     & 		 & Positive  \\
(Ourense, Spain)    & 42$^\circ$ 11' 57.0''    & CCD QHY6				& 2.48510 & L. P\'erez & 0.079\\
				    & 514		  & 						& 		  & 		 & 			\\
 \hline
Javalambre Observatory & -01$^\circ$ 00' 58.6''	  & D= 0.40 m, f= 3600 mm 				& 2.0     & R. Iglesias		 & Positive\\
(Teruel, Spain)        & 40$^\circ$ 02' 30.6''   & ProLine PL4720/e2v CCD47-20-331		& 2.68829 & J.L. Lamadrid    & 0.024\\
				       & 1957           		  & 									& 		  & N. Ma\'{i}cas    & 			\\

\hline
Aras de los Olmos Observatory   & -01$^\circ$ 06' 05.4''	 & D= 0.52 m, f= 3451 mm		& 0.5    	 & 		 & 	Positive\\
(Valencia, Spain)        		& 39$^\circ$ 56' 42.0''     & Finger Lakes Proline 16803	& 1.49875	 & V. Peris      & 0.082\\
				       			& 1280           		     & 							& 		  	 &      & 			\\	
    
 \hline
La Hita Observatory    & -03$^\circ$ 11' 09.8''  & D= 0.77 m, f= 2400 mm 		& 2.0     & N. Morales		 & Positive\\
(Toledo, Spain)        & 39$^\circ$ 34' 07.0''     & SBIG STX-16803						& 3.61480 & F. Organero      & 0.055\\
				       & 674              & 									& 		  & 			     & 			\\				    
 \hline
La Hita Observatory    & -03$^\circ$ 11' 09.8''  & D= 0.40 m, f= 1800 mm 		& 0.300    & L. Ana-Hernandez   & 	Positive \\
(Toledo, Spain)        & 39$^\circ$ 34' 07.0''    & Basler AC120				& 0.301    & F. Organero         & 0.096 \\
				       & 674           		  & 									& 		  & F. Fonseca & (Tech. problems)	\\					    
 \hline
Univ. of Athens Observatory   & 23$^\circ$ 47' 00.1'' 	 & D= 0.40 m, f= 3200 mm			& 0.5    	 & 				 & Positive\\
(Athens, Greece)        	  & 37$^\circ$ 58' 06.8''    & SBIG ST-10 CCD Camera	& 3.17352    & K. Gazeas  	 & 0.101\\
				       		  & 250           		     & 							& 		  	 & 			     & 			\\					    
 \hline
  \hline

Sant Esteve Observatory  		& 01$^\circ$ 52' 21.1''  & D= 0.40 m, f= 1600 mm		&  0.16 	 & 				 &Negative \\
(Barcelona, Spain)				& 41$^\circ$ 29' 37.5''  & CCD 1/2", Mintron 12V1C-EX & 0.16    & C. Schnabel   & 0.180 \\   							    					
								& 180					 & 							& 		  	 & 			     & 			\\
 \hline
La Sagra Observatory  		    & -02$^\circ$ 33' 55.0''   & D= 0.36 m, f= 746 mm		   &  0.300	 & 				 &	Negative\\
(Granada, Spain)				& 37$^\circ$ 58' 56.6''   & QHY174M-GPS  				   &  0.301    & N. Morales    & 0.089 \\   							    					
								& 1530					 & 							   & 		  	 & 			     & 			\\	
		\hline
	\end{tabular}}
\end{table*}

\subsection{Rotational light curve}
\label{LC_obs}

In order to obtain the rotational phase of 2002 GZ$_{32}$ at the moment of the occultation we observed the object with the 1.23-m telescope at Calar Alto Observatory in Almer\'ia (Spain) during five nights after the occultation date, three in May 2017 (23, 25 and 26) and two in June 2017 (23 and 24) obtaining a total of 128 images. The telescope and detector setups were the same as those chosen for the astrometric campaign (Table \ref{ObsSummary}). The images were acquired in 2 $\times$ 2 binning mode with an average seeing during the whole campaign of 1.9$''$ (2.1$''$ for the three nights in May and 1.6$''$ for the two nights in June) and with a Moon illumination $\sim$ 4\% during the May nights and less than 1\% during the June nights.

\begin{figure}
	\includegraphics[width=\columnwidth]{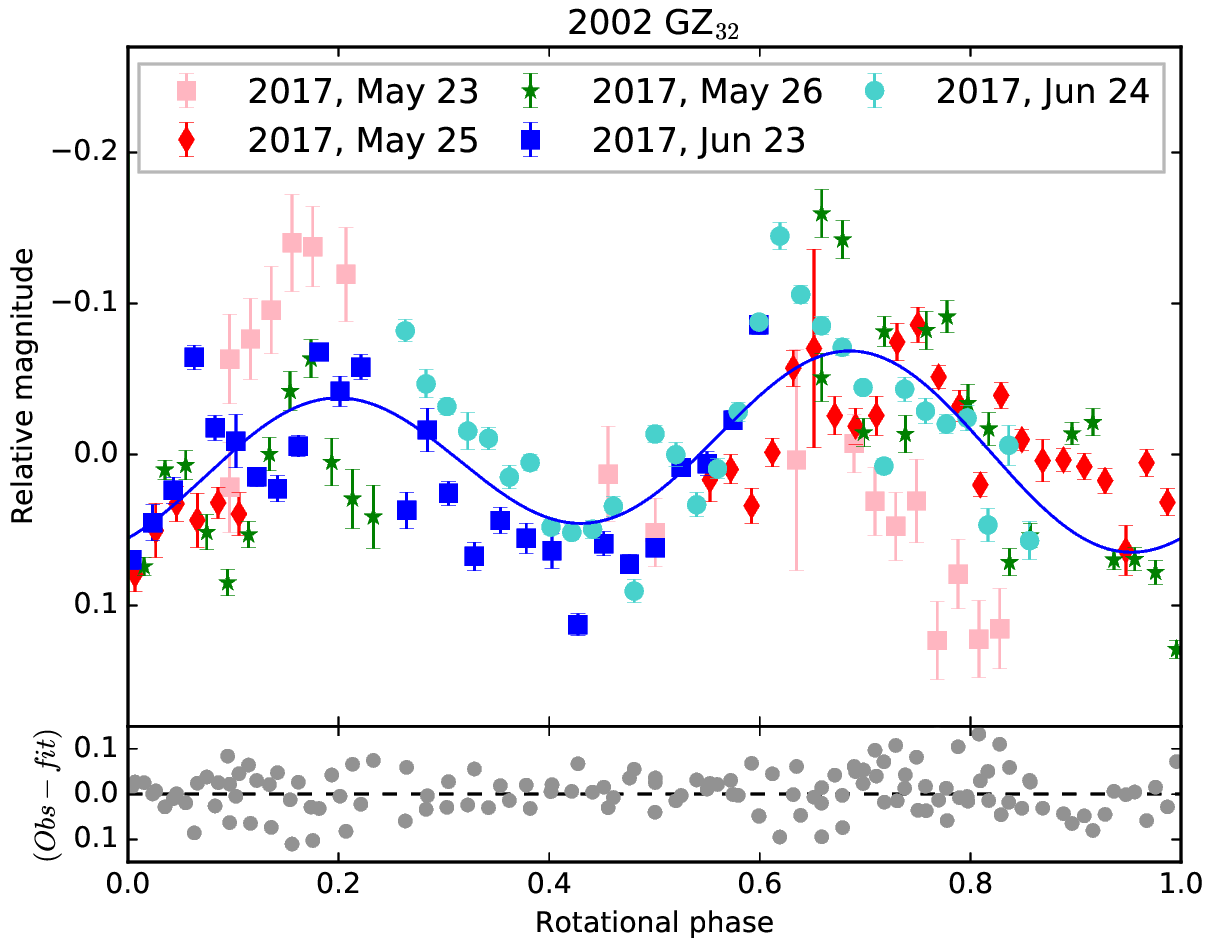}
    \caption{Rotational light curve --Relative magnitude vs. Rotational phase-- of 2002 GZ$_{32}$ obtained from data acquired few days after the occultation with the Calar Alto Observatory T1.23-m. The data have been folded with a rotational period of 5.80 $\pm$ 0.03 h \citep{Dotto2008}. The blue line represents a second-order Fourier function fit. Grey circles at the bottom panel represent the residual of the observational data to the fit. The zero rotational phase in the plot is fixed at the moment of the occultation (May 20, 2017 01:30:00 UTC).}
    \label{rotLC}
\end{figure}

Flat field and bias standard corrections were applied on the acquired science images. Aperture photometry of the object and of the selected comparison stars was extracted by means of specific routines programed in IDL. Different values for the apertures and sky ring annulus were tried for the object and (same values) for the comparison stars with the aim to minimize the dispersion of the photometry and to maximize the SNR of the object. The data processing was the same as that described in \cite{Fernandez-Valenzuela2017}. After all this process we obtained the flux of 2002 GZ$_{32}$ with respect to the comparison stars versus time --corrected for light travel times--. The rotational period of 5.80 $\pm$ 0.03 h obtained in \cite{Dotto2008} is used to fold the photometric data, obtaining the rotational light curve shown in Figure \ref{rotLC}. From this rotational light curve, acquired close to the occultation time, we determined that 2002 GZ$_{32}$ was near one of its absolute brightness minima (its minimum projected surface) during the event. Based on a fitted Fourier function, its peak-to-peak amplitude is 0.13 $\pm$ 0.01 mag, very similar to that of \cite{Dotto2008}. The zero rotational phase in the plot was chosen to be at the moment of the stellar occultation (May 20, 2017 01:30:00 UTC).

\section{Analysis of the Stellar Occultation}
\label{analysis}

Table \ref{StarDetails} summarizes the occulted star details and other related information. Our occultation data are clearly dominated by the exposure times ($\geq$ 10.60 km) rather than by the stellar diameter ($\sim$ 0.59 km at 2002 GZ$_{32}$'s distance) or Fresnel diffraction effects ($\sim$ 0.88 km). To estimate the projected diameter of the occulted star in the plane of the sky we used its B, V and K apparent magnitudes from the NOMAD catalog \citep[B = 14.900 mag, V = 14.020 mag, K = 10.290 mag,][]{Zacharias2004} and the \cite{vanBelle99} equation. The Fresnel scale value of $F = \sqrt{\lambda d/2}$ = 0.88 km results from $\lambda$ = 600 nm, the average central wavelength of the observations, and  the centaur's geocentric distance during the occultation ($d$ = 17.1033 AU).

\begin{table*}
	\centering
	\caption{Details of the occulted star and other related information.}
	\label{StarDetails}
	\begin{tabular}{l|l}   
		\hline	    
\textbf{Designation}   &  Gaia DR1 4332852996360346368, UCAC4 385-75921 (Gaia DR2 4332853000658702848)  \\
\hline
\textbf{Coordinates DR1}$^{a}$  & 	$\alpha$ = 16h 50m 08s.5375, $\delta$ = -13$^\circ$ 05' 12''.733	\\
\hline
\textbf{Coordinates DR2}$^{b}$  & 	$\alpha$ = 16h 50m 08s.5378, $\delta$ = -13$^\circ$ 05' 12''.7381	\\ 
\hline
\textbf{Magnitudes}$^{c}$    & 	B = 14.900, V = 14.020, R = 13.810, J = 11.210, H = 10.444, K = 10.290 \\
\hline
\textbf{Diameter}$^{d}$    & 	$\sim$ 0.0472 mas ($\sim$ 0.59 km at 2002 GZ$_{32}$'s distance) \\	
\hline
\textbf{Speed}$^{e}$  & 	21.19 km/s	\\
		\hline
	\end{tabular}
\begin{tablenotes}
      \small
      \item $^{a}$Gaia DR1 coordinates corrected for proper motion using the HSOY catalogue for the occultation date.
   	  \item $^{b}$Gaia DR2 coordinates corrected for proper motion for the occultation date. These coordinates became available after the occultation.
      \item $^{c}$From the NOMAD catalog \citep{Zacharias2004}.
	  \item $^{d}$Estimated from the B, V, K magnitudes using the \cite{vanBelle99} equation.
	  \item $^{e}$Speed of 2002 GZ$_{32}$ with respect to the star as seen from Earth.
\end{tablenotes}
\end{table*}

\begin{figure}
	\includegraphics[width=\columnwidth]{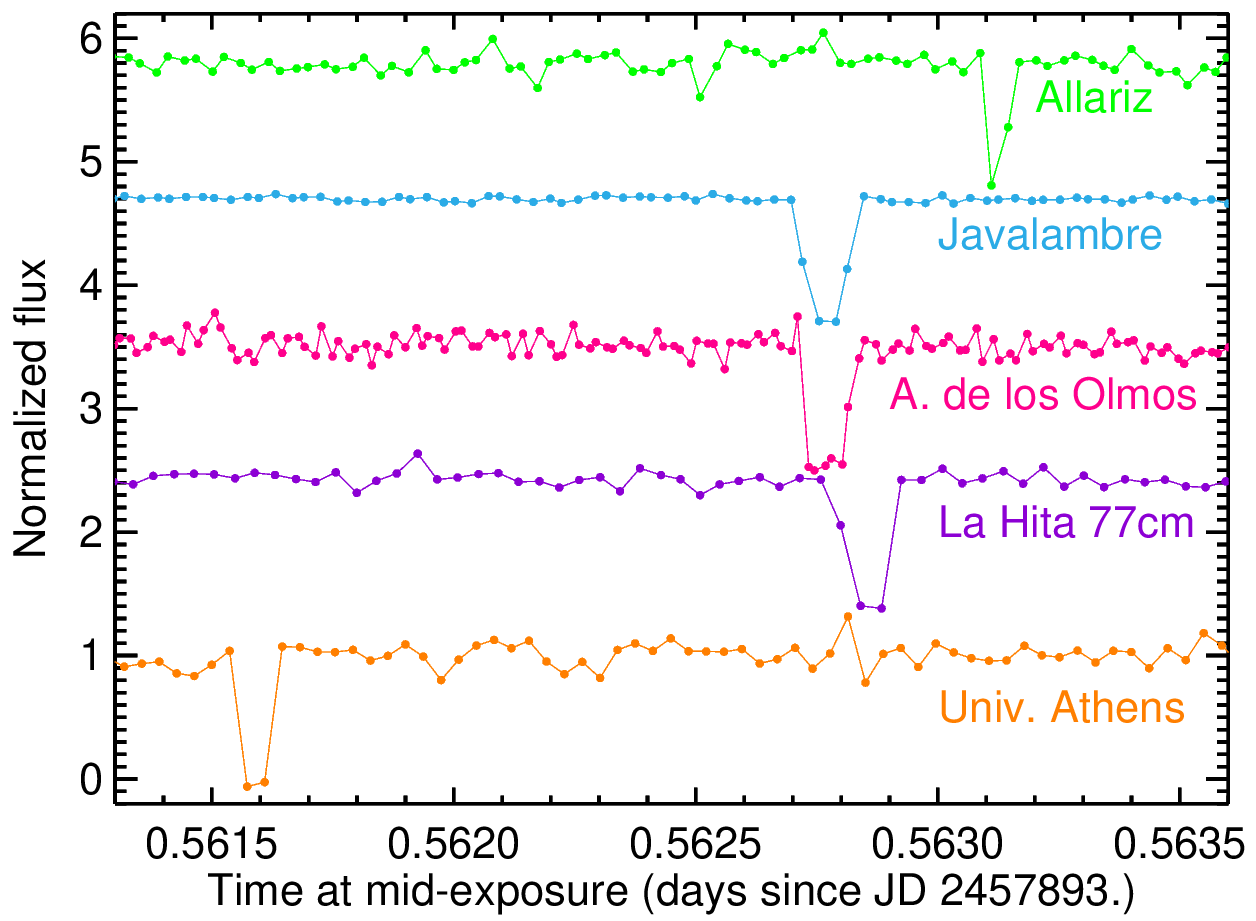}
    \caption{Stellar occultation light curves (normalized flux versus time at mid-exposure) from the five sites where the occultation was recorded. The light curves are ordered from the northern (top) to the southern (bottom) sites: Allariz, Javalambre, Aras de los Olmos, La Hita (Spain), University of Athens (Greece). The light curves have been shifted in flux for a better viewing. Error bars are not shown in order to avoid a messy plot. The standard deviation of the measurements are indicated in Table \ref{ObservDetails}.}
    \label{OccLCs}
\end{figure}

The occultation light curves (see Figure \ref{OccLCs}) are used to derive the times of disappearance and reappearance of the star behind the 2002 GZ$_{32}$ limb. These `ingress' and `egress' times determine different segments in the plane of the sky for each observing site, those segments (expressed in time) can be directly translated to distances using the apparent motion of 2002 GZ$_{32}$ relative to the occulted star which is 21.19 km/s in our case (see Table \ref{StarDetails}). The obtained segments in the plane of the sky are called `chords'. These chords are used to achieve the results described in Section \ref{results}. The ingress and egress times (and associated uncertainties) at each site are obtained by means of the creation of a synthetic light curve using a square-well model convolved with the star apparent size, the Fresnel scale and the exposure time (the star apparent size and the Fresnel effects are negligible in our case, when compared with the exposure times). This synthetic light curve is compared with the data and the difference is iteratively minimized using a $\chi^2$ metric, as is described in \cite{Ortiz2017} and in \cite{Benedetti-Rossi2016} --and references therein--. The times of disappearance and reappearance obtained are shown in Table \ref{Chords}, from these ingress/egress times six chords were obtained with values (in km) also included in Table \ref{Chords}. The uncertainties in the times were small, except for the University of Athens observations due to the long integration time and large deadtimes of the detector. We also note that the video from the La Hita Observatory T-40cm was only used to confirm the positive detection from the same site using the T-77cm and not to perform the limb
fit (see Section \ref{properties}), which is hence based on five chords, not six. The reason is two-fold: an estimated 1\% of the frames
during the video acquisition were lost and the absolute timing was offset by an unknown amount due to a computer problem. 

\begin{table*}
	\centering
	\caption{Ingress and egress times derived from the occultation light curves. May 20$^{\rm th}$ 2017. The results obtained from La Hita 40-cm telescope (in parenthesis in the table) were not used to perform the limb fit due to technical problems which prevented a good timing during the observation, in any case, the size of the chord is used as a consistency check of the other chord obtained from the same site with the 77-cm telescope, both chords are consistent within error bars.}
	\label{Chords}
	\begin{tabular}{ccccc}   
		\hline	
		\hline	    
\textbf{Observatory, Country}  &  \textbf{Ingress (UT)} & \textbf{Egress (UT)} & \textbf{Chord size (km)}  & \textbf{Shift (s)} \\
\hline
\textbf{Allariz, Spain}  &	1:30:51.50 $\pm$ 0.50 &  1:30:55.75 $\pm$ 0.15 & 90.1 $\pm$ 13.8 & +0.013\\
\textbf{Javalambre, Spain}	 & 1:30:19.00 $\pm$ 0.10 & 1:30:27.15 $\pm$ 0.08 & 172.7 $\pm$ 3.8 & +0.006\\	
\textbf{Aras de los Olmos, Spain} & 1:30:19.50 $\pm$ 1.70 & 1:30:28.25 $\pm$ 0.14 & 185.4 $\pm$ 39.0 &-0.771\\	
\textbf{La Hita 77-cm, Spain} & 1:30:26.16 $\pm$ 0.23 & 1:30:34.19 $\pm$ 1.70 & 170.2 $\pm$ 40.9 & -0.258\\
( La Hita 40-cm, Spain & 1:30:25.86 $\pm$ 0.90 & 1:30:34.32 $\pm$ 0.69 & 179.3 $\pm$ 33.7 ) & \\
\textbf{Univ. of Athens, Greece} & 1:28:37.02 $\pm$ 2.66 & 1:28:43.31 $\pm$ 2.26 & 133.3 $\pm$ 104.3 & +2.048\\
		\hline
	\end{tabular}
\end{table*}

\begin{figure}
	\includegraphics[width=\columnwidth]{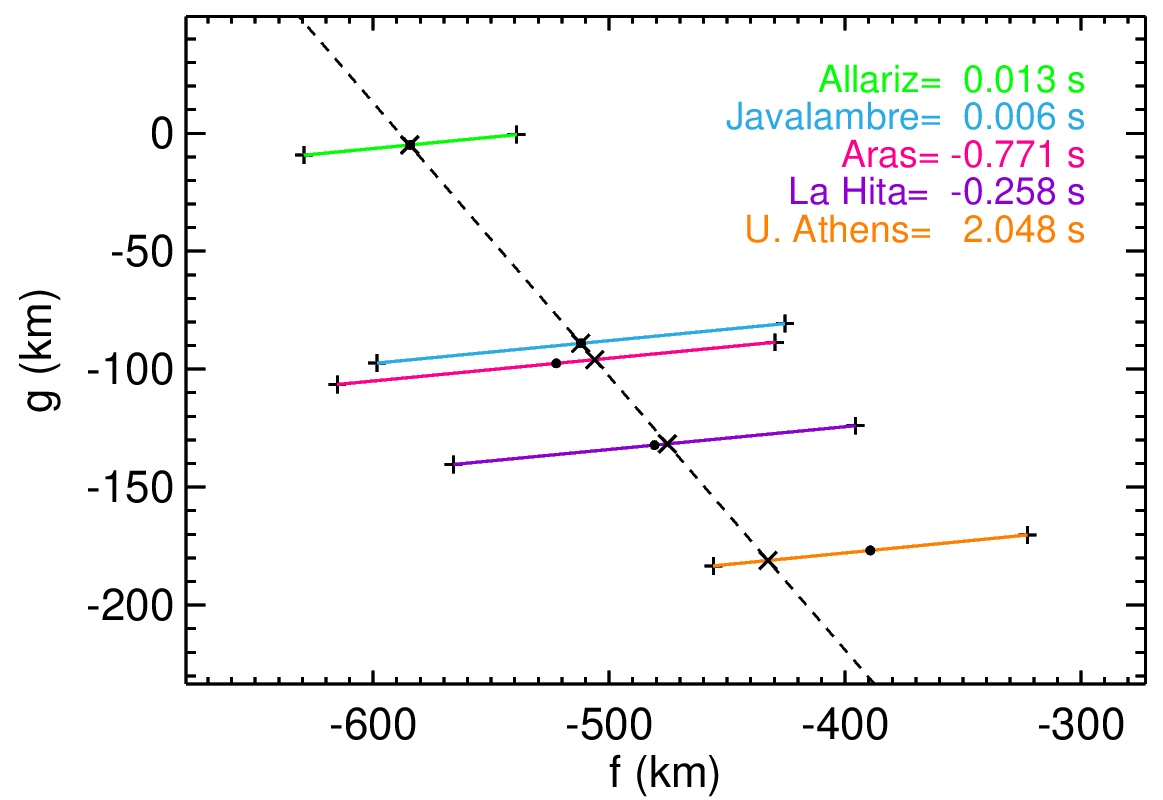}
    \caption{Chords in the sky plane and time shifts (in seconds) applied to the occultation chords. To calculate these shifts we performed a linear fit (dashed line) to the centers of the chords (black dots) obtaining the new centers (black crosses). The obtained shifts are below the mean uncertainties of each chord, as is shown in Table \ref{Chords}. The f and g axes are centered in the center obtained from the elliptical fit ($f_{\rm c}$ , $g_{\rm c}$) = (-501 km, -107 km), as in Figure \ref{elliptical_fit}. Note that ($f_{\rm c}$ , $g_{\rm c}$) provides the offsets with respect to positions obtained using the Jet Propulsion Laboratory orbit JPL\#23, assuming that the Gaia DR2 occulted star position is correct.}
    \label{ChordsShifts}
\end{figure}



\section{Results}
\label{results}

Using the chords obtained as described in Section \ref{analysis} it is possible to derive information about the physical properties of the object itself (Sections \ref{properties} and \ref{3dshapes}) and analyze the enviroment around 2002 GZ$_{32}$ searching for possible rings or debris material orbiting this centaur (Section \ref{rings}).

\subsection{Limb fit, diameter and albedo of 2002 GZ$_{32}$}
\label{properties}

From the five chords in bold in Table \ref{Chords} we have ten chord extremities that can be used to obtain the projected shape of 2002 GZ$_{32}$ at the moment of the occultation (i.e., the instantaneous limb). We fitted an ellipse and determined its parameters, as usually done in previous occultation papers \citep[see e.g.][]{Souami2020}, by minimizing the $\chi^2$ function using the simplex method adapted in programs and routines from Numerical Recipes \citep{Press1992}. In our case, $\chi^2 = \sum_{i=1}^{10}(r_{i,\rm obs}-r_{i,\rm com})^2 / \sigma^{2}_{i,\rm r} $, where r is the radius from the center of the ellipse ($f_{\rm c}$ , $g_{\rm c}$), the subscript ``com'' means computed, the subscript ``obs'' means observed, and $\sigma_{i,\rm r}$ are the uncertainties on the determination of the extremities. It is important to note that the center of the fitted ellipse ($f_{\rm c}$ , $g_{\rm c}$) provides the offsets in the celestial coordinates (R.A., Decl.) to be applied to the adopted ephemeris, JPL\#23, assuming that the Gaia DR2 occulted star position is correct. From this fit we obtained the center of the ellipse ($f_{\rm c}$ , $g_{\rm c}$), the projected or apparent semi-axes of the ellipse (a', b') and the position angle of the minor axis of the ellipse (P') --which is the apparent position angle of the pole measured eastward from celestial north--. Given that the ellipse fit is not linear in the parameters an estimate of the uncertainty of the parameters using a $\chi^2$ statistics can result in an underestimation of errors. For this reason, the uncertainties in the five parameters of the fitted ellipse were estimated by means of a Monte-Carlo method. We ramdomly generated the ten chord extremities $10^6$ times, meeting the requirements provided by their corresponding error bars. Then, we obtained the best-fitted ellipse (in terms of a $\chi^2$ minimization method as explained above) for each one of these sets of randomly generated extremities. Finally, we obtained a distribution of $10^6$ values for each ellipse parameter and computed the corresponding standard deviations to provide our 1-$\sigma$ error estimates (see errors in a', b', $f_{\rm c}$, $g_{\rm c}$ and P' in Table $\ref{limbfits}$).

\begin{figure}
	\includegraphics[width=\columnwidth]{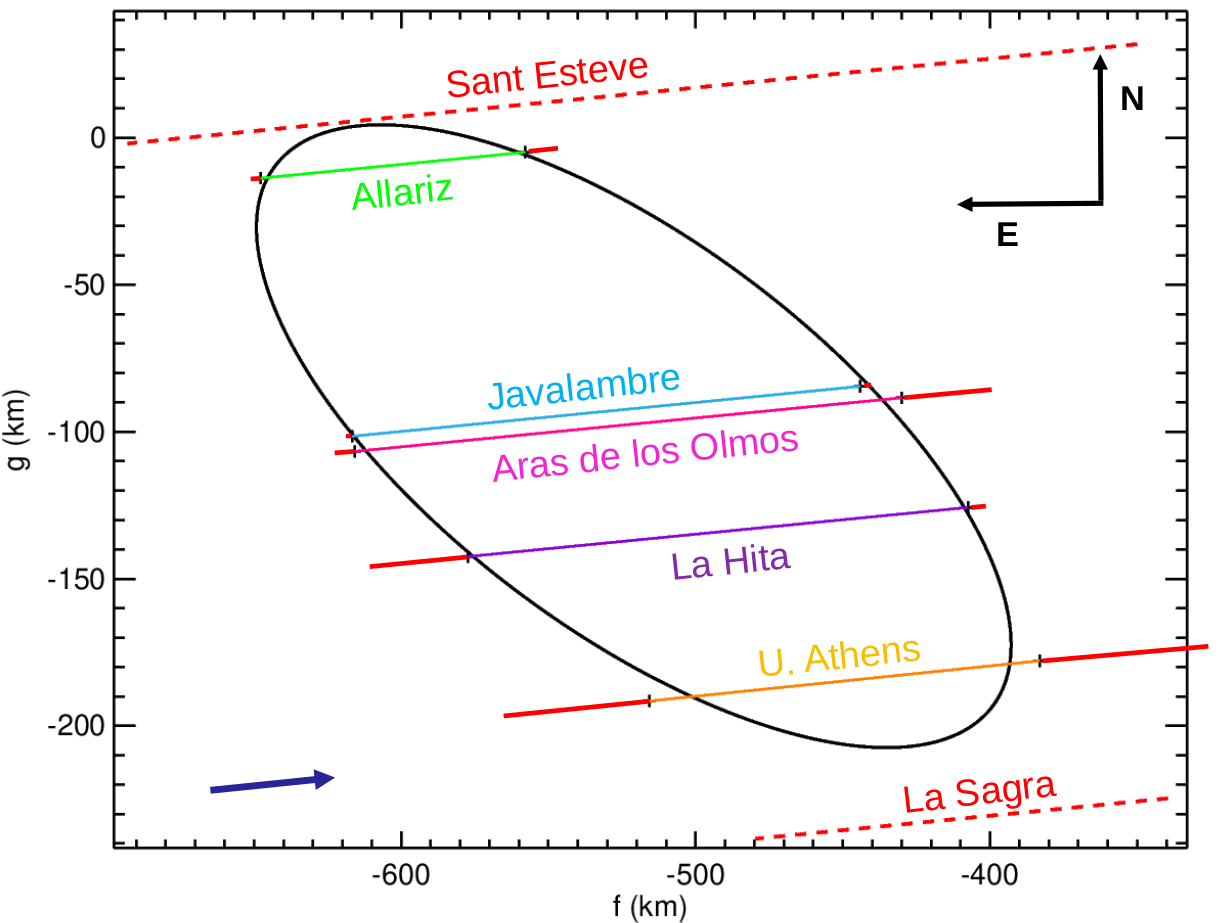}
    \caption{Elliptical fit to the five occultation shifted chords obtained from the light curves shown in Figure \ref{OccLCs}. This fit determines the limb of 2002 GZ$_{32}$ at the moment of the occultation. The best fitted ellipse has major axes of 305 $\pm$ 17 km $\times$ 146 $\pm$ 8 km. The red solid lines in the extremities of the chords are the 1$\sigma$ uncertainties of the ingress/egress times. The blue arrow shows the direction of the shadow motion. The lack of detections at Sant Esteve (Northern red dashed line) and La Sagra Observatory (Southern red dashed line) constrains the limb fit considerably. The f and g axes are centered in the center obtained from the elliptical fit (-501 km, -107 km).}
    \label{elliptical_fit}
\end{figure}

The Athens chord had a large time uncertainty and led to a bad fit ($\chi^2$ = 29.5), so we decided to slightly shift the chords performing an error bar-weighted linear fit to the centers of the chords (see Figure \ref{ChordsShifts}) under the assumption that the projected body is an ellipse and the centers of its chords must be aligned. Note that an ellipse is a plausible projected shape for 2002 GZ$_{32}$ taking into account that an icy body with its size is expected to be in -or near to- hydrostatic equilibrium. This time shift approach has been used and justified for other stellar occultations by TNOs and Centaurs, like e.g. was done for the TNO Quaoar in \cite{Braga-Ribas2013} and it is also supported by the errors in time (of the order of several tenths of a second and even larger) reported when using NTP + PC based time for different operating systems and camera control software \citep[e.g.][]{Barry2015,Ortiz2020b}. The shifts in time obtained from this linear fit are, in all cases, below the chord error bars, as is shown in Table \ref{Chords} (of course, these shifts in time can be directly translated to shifts in kilometers using the speed of 2002 GZ$_{32}$ with respect to the star as seen from Earth --see Table \ref{StarDetails}--). It is important to note that the ellipse obtained from the unshifted chords fit is fully compatible, within the error bars, with the ellipse obtained from the shifted chords fit. In particular, the semiminor axes obtained from both elliptical fits are the same, and the semimajor axes differ by less than 5\%. The limb fit solutions obtained using the original chords and the shifted chords are shown in Table \ref{limbfits}. As both solutions are similar and consistent within error bars, we choose the one with the smallest $\chi^2$ as our best solution. The axis ratio of the preferred ellipse is large (a'/b' = 2.09 $\pm$ 0.24) which indicates that 2002 GZ$_{32}$ is a very elongated body, as can be seen in Figure \ref{elliptical_fit} where the 5 (shifted) chords obtained from the occultation are plotted in the plane of the sky with their associated error bars and the best-fitted ellipse to their extremities. The two negative chords closest to the fitted limb (from Sant Esteve and La Sagra in Spain) are also plotted in Figure \ref{elliptical_fit}. 

\begin{table}
	\centering
	\caption{Elliptical fits obtained for the original and shifted chords. The area-equivalent diameter ($D_{\rm eq}$ ) and geometric albedo ($p_{\rm V}$) derived from these solutions are also included.}
	\label{limbfits}
	\resizebox{\columnwidth}{!}{%
	\begin{tabular}{c|cc}   
		\hline	    
   &  \textbf{Original chords} & \textbf{Shifted chords}\\
\hline
\textbf{(a', b')} &(146.0 $\pm$ 12.0 km, 73.0 $\pm$ 4.6 km) & (152.6 $\pm$ 8.4 km, 73.0 $\pm$ 4.2 km)\\
\hline
\textbf{($f_{\rm c}$, $g_{\rm c}$)} & (-500 $\pm$ 9 km, -112 $\pm$ 6 km) & (-501 $\pm$ 6 km, -107 $\pm$ 5 km) \\
\hline
\textbf{P'} & 318$^\circ$ $\pm$  3$^\circ$ & 322$^\circ$ $\pm$  3$^\circ$\\
\hline
\textbf{$\chi^2$} & 29.5 & 4.0 \\
\hline
\hline
\textbf{$D_{\rm eq}$} & 206 $\pm$ 15 km & 211 $\pm$ 12 km \\
\hline
\textbf{$p_{\rm V}$} & 0.044 $\pm$ 0.009 & 0.043 $\pm$ 0.007 \\
		\hline
	\end{tabular}}
\end{table}

Using the values of the axis instead of the semi-axis of the best-fitted ellipse we have (2a', 2b') = (305 $\pm$ 17 km, 146 $\pm$ 8 km), from where we can derive the area-equivalent diameter of 2002 GZ$_{32}$ at the moment of the occultation that turns out to be $D_{\rm eq_{min}}$ = 211 $\pm$ 12 km. This area-equivalent diameter is only a lower limit of the real equivalent diameter of the object because, as was stated in Section \ref{LC_obs}, we know that 2002 GZ$_{32}$ was very close to its absolute brightness minimum at the instant of the occultation, which means that its projected area was also very close to its minimum.

From the best-fitted limb obtained we derived the geometric albedo at V-band of the surface of 2002 GZ$_{32}$ at the moment of the occultation using the expression: $p_{\rm V}$ = 10$^{0.4(V_{\rm sun}-H_{\rm V})}/(A/\pi$), where $V_{\rm sun}$ is the V-magnitude of the Sun ($m_{\rm V}$ = -26.74 mag), $H_{\rm V}$ is the absolute magnitude of 2002 GZ$_{32}$ at V-band ($H_{\rm V} = 7.39 \pm 0.06$ mag, \cite{Alvarez-Candal2019}), and $A$ is the projected area of the centaur that can be directly obtained from the occultation elliptical fit. Finally, to obtain the right albedo, we have to correct $H_{\rm V}$ taking into account the rotational phase at the instant of the occultation. As we know that the object was close to its minimum projected area during the occultation (Section \ref{LC_obs}) we should add $\Delta \rm m/2$ = 0.13/2 mag = 0.065 mag to $H_{\rm V}$ to derive the absolute magnitude at the moment of the occultation. A more precise value to be added to $H_{\rm V}$ during the occultation can be obtained from the Fourier fit to the light curve, and it turns out to be of 0.055 mag. After this correction to obtain the instantaneous absolute magnitude, the geometric albedo calculated was $p_{\rm V}$ = 0.043 $\pm$ 0.007. This value is slightly larger than the geometric albedo derived from the radiometric method using Herschel, Spitzer and ALMA thermal data \citep[$p_{\rm V}$ = 0.037 $\pm$ 0.004 and $p_{\rm V}$ = 0.036$^{+0.006}_{-0.005}$ from][respectively]{Duffard2014,Lellouch2017}, but still compatible within error bars.

\subsection{Possible 3-D shapes of 2002 GZ$_{32}$}
\label{3dshapes}

Under the assumption that the rotational light curve of 2002 GZ$_{32}$ is produced by a rotating triaxial ellipsoid with semi-axes a, b and c (a > b > c, with c the axis of rotation) in an equator-on geometry (aspect angle = 90$^\circ$), it is possible to relate the amplitude of the light curve ($\Delta \rm m$ = 0.13 mag) with the axes ratio a/b, using the expression $\Delta \rm m = 2.5 \cdot log(a/b)$ from which we obtained a/b = 1.13. As the best-fit ellipse obtained from the occultation is very close to the minimum projected area, it is clear that b $\simeq$ a', and then, the longest semi-axis (a) can be obtained from the amplitude-derived a/b value (a = 172.4 km). For the shortest axis (c) it is only possible to provide an upper limit, because the pole position of 2002 GZ$_{32}$ is unknown (i.e., the aspect angle is unknown). Then, a possible 3-D model for 2002 GZ$_{32}$ would have the axes: 2a = 345 km, 2b= 305 km, and 2c = 146 km. From this 3-D shape we can derive a maximum area-equivalent diameter for the centaur (because we only have an upper limit for semi-axis c) obtaining $D_{\rm eq_{max}}$ = 225 $\pm$ 12 km. A `mean' area-equivalent diameter of <$D_{\rm eq}$> = 218 $\pm$ 12 km is obtained from the minimum ($D_{\rm eq_{min}}$) and maximum ($D_{\rm eq_{max}}$) area-equivalent diameters. This mean diameter is smaller, and not in agreement with the radiometric diameter obtained from thermal data using Herschel and Spitzer \citep[$D = 237 \pm 8$ km;][]{Duffard2014},  but it is still compatible within error bars with the latest radiometric estimate adding ALMA measurements to Herschel/Spitzer data by \cite{Lellouch2017}: $D = 237^{+12}_{-11}$ km.

The aspect angle, however, does not have to be 90$^\circ$ (2c $\le$ 146 km) and, as the occultation took place very close to the minimum of the rotational light curve (minimum projected area), it is possible to search for the aspect angles and triaxial ellipsoids compatible with the projected axial ratio obtained from the occultation (2.09 $\pm$ 0.24) and the peak-to-peak rotational light curve amplitude (0.13 $\pm$ 0.01 mag). To do this, we carried out a $\chi^2$ minimization of the results obtained from a grid search for the aspect angle and the axes (2a, 2b, 2c) compatible with the aforementioned axial ratio and $\Delta$m. The aspect angle was explored from 0$^\circ$ to 90$^\circ$ at intervals of 0.5$^\circ$ and the axes were explored from (294 km, 294 km, 120 km) to (580 km, 316 km, 146 km) at steps of 4 km. This choice ensures that the space of possible 3D solutions is sampled densely enough and that the minimum $\chi^2$ solution lies within those intervals. The $\chi^2$ function to be minimized was defined as $\chi^2 = (\Delta {\rm m} - 0.13)^2/0.01^2 + (A_{\rm ratio} - 2.09)^2/0.24^2 + (b - 152.6)/8.4^2$, with $\Delta \rm m$ the light curve amplitude, $A_{\rm ratio}$ the projected axial ratio, and $b$ the semiaxis $b$ obtained for each triaxial ellipsoid generated during the grid search. From this search we obtained a family of possible solutions of triaxial ellipsoids and aspect angles. The axes (2a, 2b, 2c) that meet the constraints ranged from (330 km, 294 km, 120 km) to (402 km, 314 km, 144 km), for aspect angles from 71.5$^\circ$ to 90.0$^\circ$. The triaxial model that minimizes the $\chi^2$ has axes 2a = 366 km, 2b = 306 km, 2c = 120 km, with an aspect angle of 76$^\circ$.

\begin{figure}
	\includegraphics[width=\columnwidth]{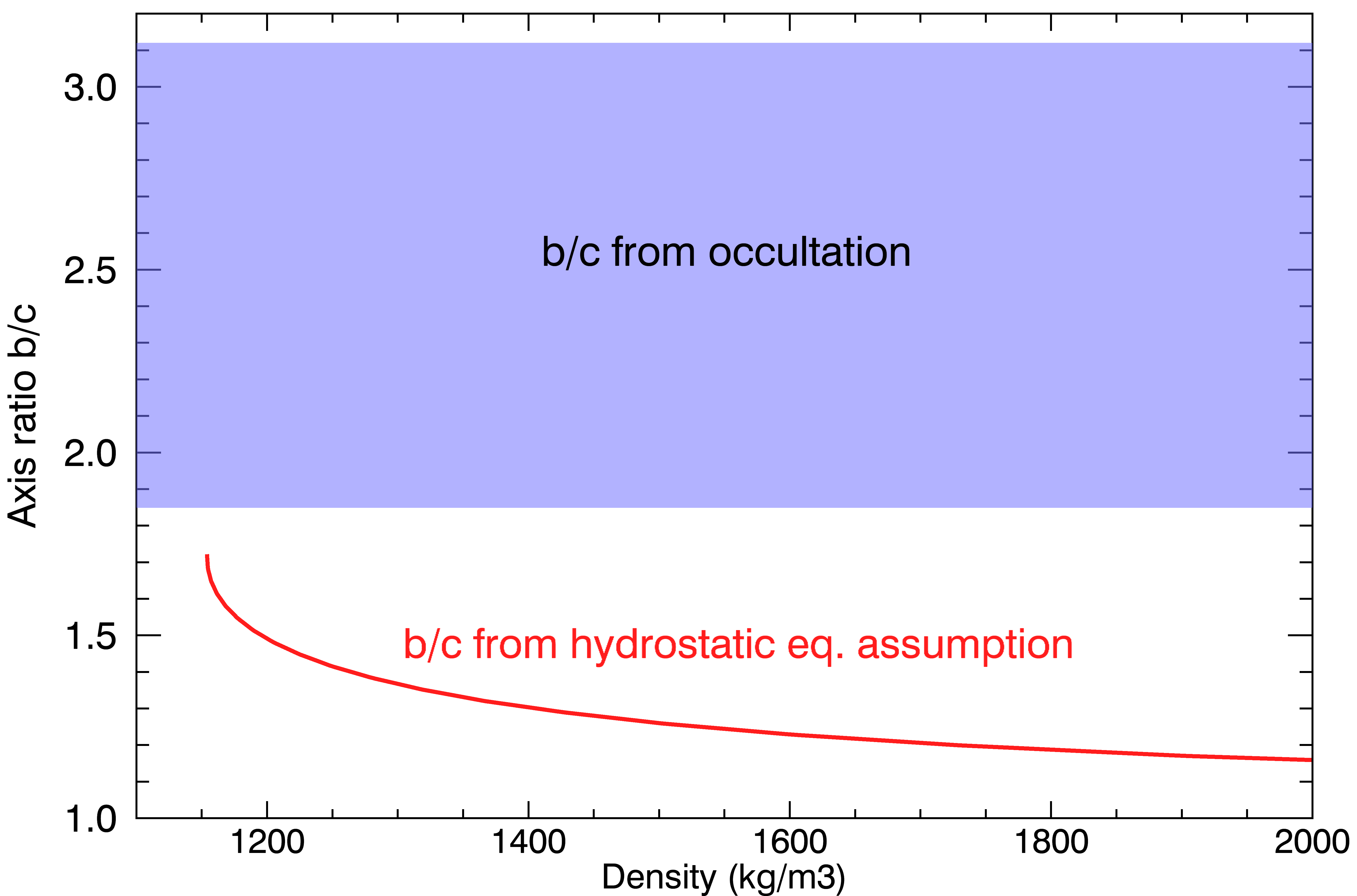}
    \caption{Axis ratio b/c versus density. The red curve corresponds to the possible b/c values for different densities for a triaxial body in hydrostatic equilibrium rotating at 5.80 h, following the Chandrasekhar (1987) formalism. The violet band represents the range of possible b/c values permited by the occultation results. The lower limit of the band is obtained from the instantaneous limb derived from the occultation (b/c = 2.09 for an aspect angle of 90$^\circ$) minus the corresponding 1$\sigma$ error bar. The upper limit of the band is obtained from the preferred triaxial solution --in terms of a $\chi^2$ minimization-- (b/c = 2.55) plus the corresponding 1$\sigma$ uncertainty. As the red curve is not crossing the violet band, the shape obtained from the occultation is not compatible with a triaxial body in hydrostatic equilibrium.}
    \label{bc_vs_density}
\end{figure}

We know that 2002 GZ$_{32}$ produces a double peaked rotational light curve. The most natural explanation is that the object must be triaxial in shape. If it is a triaxial body in hydrostatic equilibrium, the minimum density that it could have in order to produce a rotational amplitude of 0.13 mag (a/b = 1.13) rotating at 5.80 h, would be 1161 kg/m$^3$ using the \cite{Chandrasekhar1987} formalism. If the aspect angle is smaller than 90$^\circ$, the a/b ratio would be larger than 1.13 and the density would be higher too. But the axial ratio b/c corresponding to a hydrostatic equilibrium body with density 1161 kg/m$^3$ rotating in 5.80 h would be 1.61, whereas the minimum value from the occultation is 2.09. It could be even larger if the object's aspect angle is smaller than 90$^\circ$. Hence, the occultation results seem to be incompatible with a hydrostatic equilibrium shape, as is illustrated in Figure \ref{bc_vs_density}.

On the other hand, one must note that 1161 kg/m$^3$ is a large density for an object of 2002 GZ$_{32}$'s size, given what we know from several TNOs and centaurs of different sizes, that the density decreases with size \citep[e.g.,][]{Grundy2015,Bierson2019}. It could be, as it was the case for Haumea \citep{Ortiz2017}, that the real density of the body is considerably lower than the one derived from the hydrostatic equilibrium assumption for a homogeneous body. In the case of Haumea, the real density is in the 1900 to 2000 kg/m$^3$ range whereas the hydrostatic equilibrium value was at least 2500 kg/m$^3$. Something similar could be happening with 2002 GZ$_{32}$. Contrary to Haumea, 2002 GZ$_{32}$ appears to be too small to be differentiated, so this does not appear to be a plausible explanation.

\subsection{Search for rings or debris around 2002 GZ$_{32}$}
\label{rings}

\begin{figure*}
	\includegraphics[width=\textwidth]{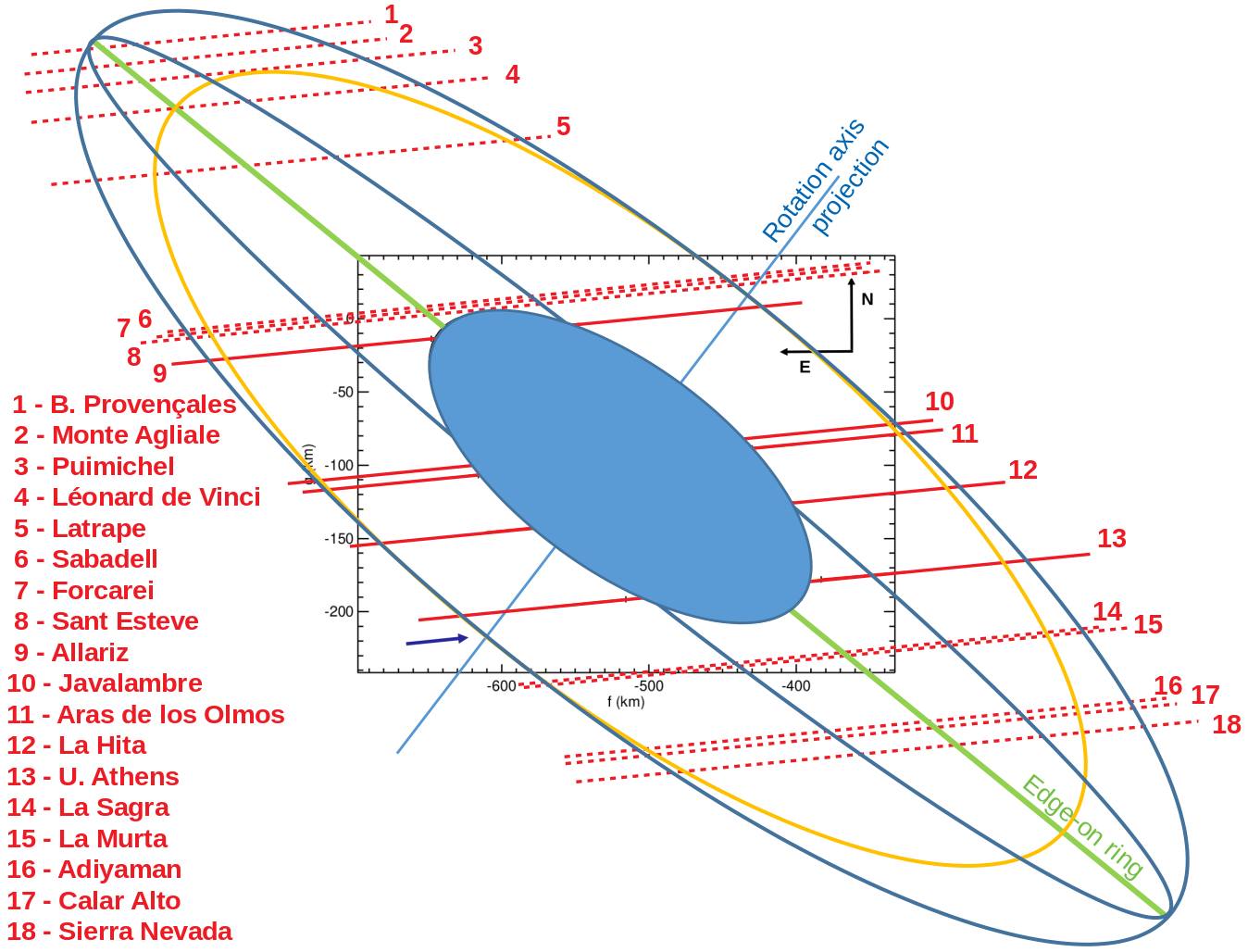}
    \caption{Scheme showing a few potential ring geometries around 2002 GZ$_{32}$, assuming the centaur is rotating around its minor axis and assuming that the rings are equatorial. Rings with different radii and pole orientations are shown, from pole angle of $\sim$ 90$^\circ$ (edge-on green ring) to pole angles of $\sim$ 84$^\circ$ and $\sim$ 72.5$^\circ$ (blue rings) and $\sim$ 66.5$^\circ$ (yellow ellipse). The plot illustrates that all ring systems with tilt angles smaller than $\sim$ 84$^\circ$ should have given rise to detectable signals in the positive occultation chords (solid red lines) that sampled the main body, provided that the potential rings had sufficiently high optical depths. Potential rings with tilt angles larger than $\sim$ 84$^\circ$ would still be detectable in the light curves that did not detect the primary body (dashed red lines). However, no ring has been detected. Upper limits to the optical depths and widths of the potential rings are shown in Table \ref{ringconstraints}}
    \label{ring_geometries}
\end{figure*}

No secondary drops below the 3$\sigma$ level were detected in the positive occultation light curves (Figure \ref{OccLCs}), nor in the `best' (in terms of flux dispersion) of the negative light curves. We can constrain the presence of ring/debris material using those `best light curves' among the positive and negative detections. Within the positive detections, the Javalambre Observatory light curve is the one with the smallest flux dispersion ($\sigma_{\rm flux}$ = 0.024). Given that the exposure time at Javalambre was 2 s, and the speed of the star with respect to the observer was 21.2 km/s, in 2 s we would see a decrease in flux of 100\% if a ring of 42.4 km in width and 100\% opacity had been present. Of course we can rule this out, but the 3$\sigma$ limit implies that a maximum flux drop of 7.2\% could have been missed, which implies that a ring smaller than $\sim$ 3 km in width would have been missed if the opacity of the ring were 100\%. For an intermediate opacity ring, such as that of Haumea or Chariklo, a ring smaller than 6 km in width would have been missed. Many combinations of widths and opacities would be possible. The minimum opacity would be for an optical depth of $\tau_{\rm min}$ = 0.07 and a ring width of 42.4 km. The best constraint to a putative ring within the positive observations was obtained from the La Hita 40-cm telescope data, with an integration time of only 0.3 s without deadtimes ($\sigma_{\rm flux}$ = 0.096). From these data, rings with widths larger than 3.7 km and 1.8 km could have been detected at 3$\sigma$ for opacities of 50\% and 100\%, respectively. In this case, the minimum opacity would be for an optical depth of $\tau_{\rm min}$ = 0.34 and a ring width of 6.4 km. Note that, if we assume that 2002 GZ$_{32}$ is rotating around its minor axis, an equatorial ring system (if any) would be expected (i.e., the ring pole would be aligned with the body's rotation axis). Then, the Javalambre and La Hita T40-cm data constrain rings for a good fraction of geometries around 2002 GZ$_{32}$ (see Figure \ref{ring_geometries}), but wider or optically thinner rings for a larger number of ring system geometries cannot be ruled out.

\begin{figure}
	\includegraphics[width=\columnwidth]{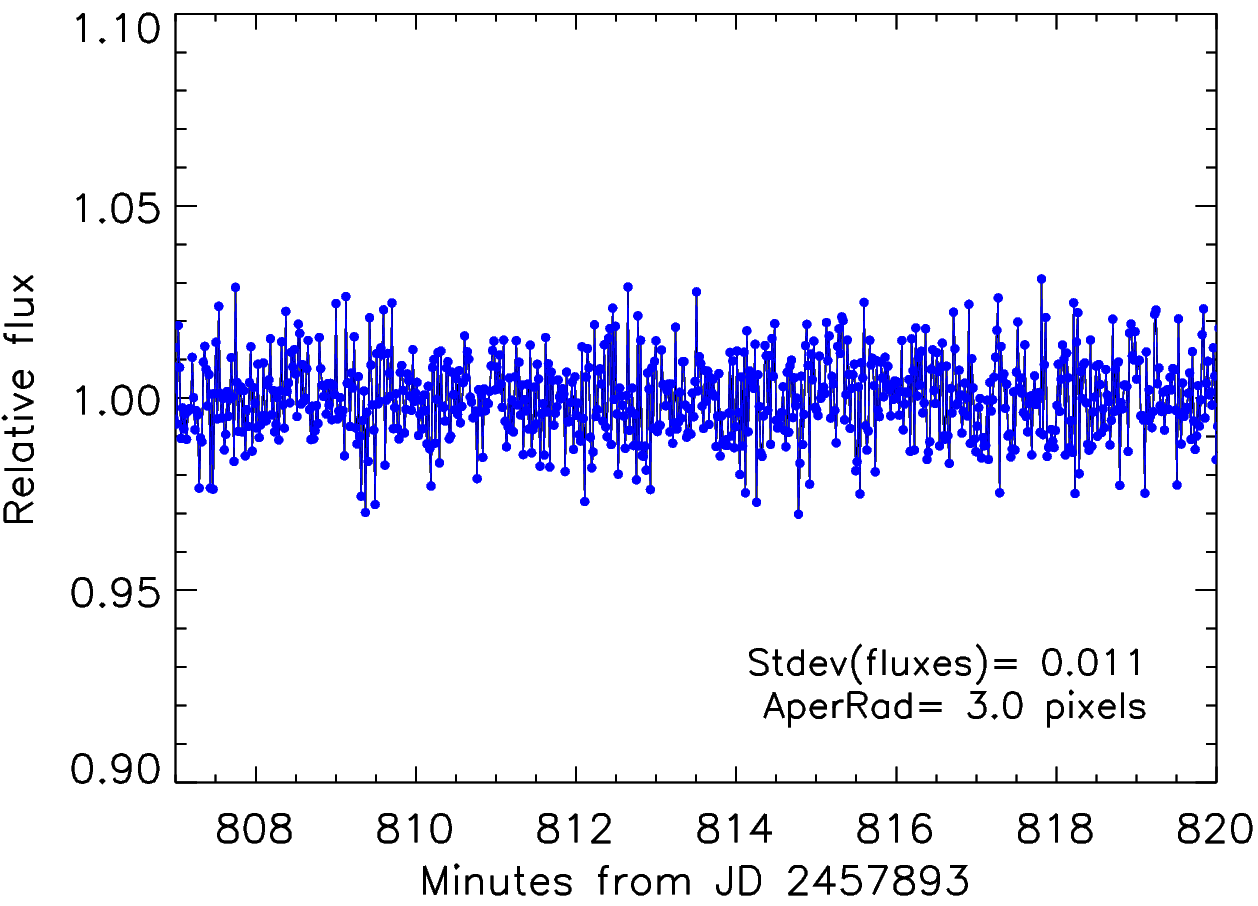}
    \caption{The negative light curve obtained from the 1.23-m telescope at Calar Alto Observatory (Almer\'ia, Spain). This is the best light curve, in terms of flux dispersion, closest to the shadow limb of 2002 GZ$_{32}$. Calar Alto is located at a projected distance on the sky plane of 307 km south from the center of the limb fit, following the ellipse major axis direction (i.e. the direction of the major axis of an equatorial ring).}
    \label{CAHA_GZ32_LC}
\end{figure}

Another possibility could be that the ring or debris material would be in a near to edge-on geometry, in this case the ring would be undetectable in the positive occultation light curves, but it could be detected in the negative light curves closest to the body's limb  (e.g., Sant Esteve, La Sagra, see Table \ref{ObservDetails} and Figure \ref{elliptical_fit}; Calar Alto, see Table \ref{SummaryObserv}), as schematically shown in Figure \ref{ring_geometries}. The latter is a reasonable assumption for an equatorial ring around 2002 GZ$_{32}$, taking into account the large apparent oblateness of the projected shape derived from the occultation ($\Psi \sim$ 71.5$^\circ$-90.0$^\circ$). The best light curve (in terms of flux dispersion) closest to the object's limb was obtained with the 1.23-m telescope at Calar Alto Observatory, Almeria, Spain ($\sigma_{\rm flux}$ = 0.011, see Figure \ref{CAHA_GZ32_LC}) with exposure times of 0.778 s. The Calar Alto negative detection is at a projected distance on the sky plane of 307 km south from the center of the limb fit ($f_{\rm c}$, $g_{\rm c}$), following the ellipse major axis direction (which would correspond to the direction of the major axis of an equatorial ring). In other words, the Calar Alto chord intersects the extended major axis at 307 km distance to the center of the limb fit (note that this distance is not the distance to the centerline which is shown in Table \ref{SummaryObserv}). This distance is close enough to the object to allow the detection of an edge-on ring. As a reference, the two rings around the centaur Chariklo, with a size slighly larger than 2002 GZ$_{32}$, have orbital radii of 391 km and 405 km, respectively \citep{Braga-Ribas2014}. From this light curve we could have detected a ring around 2002 GZ$_{32}$ at 3$\sigma$ if it had a width $>$ 0.5 km, for an opacity of 100\%. For an opacity of 50\% we could have detected rings with widths $>$ 1.1 km at the 3$\sigma$ level of noise (see Table \ref{ringconstraints}).

\begin{table*}
	\centering
	\caption{Ring constraints obtained from telescopes > 0.8 m located at different sites north or south of the limb. The second and third columns are the ring width (w) and minimum optical depth ($\tau_{\rm min}$) for a 3$\sigma$ detection. The fourth and fifth columns are the ring widths detectable at 3$\sigma$ for opacities of 50\% and 100\% (completely opaque ring), respectively. Sites from north to south sorted by projected distance on the sky plane to the reconstructed centerline, following the ellipse major axis direction, which would correspond to the direction of the major axis of an equatorial ring (D$_{\rm Major Axis}$). Note that these distances are different from those shown in Table \ref{SummaryObserv}, which are the minimum distances from the center of the limb fit to the different chords, i.e. following chord perpendicular direction. Positive distances indicate sites north of the centerline, negative distances indicate sites south of the centerline. The Roche limit is around 500-560 km, for densities 700-1000 kg/m$^3$, respectively. The existence of rings at distances larger than the Roche limit is very unlikely.}
	\label{ringconstraints}
	{
	\begin{tabular}{c|cccccc}   
		\hline
   \textbf{Observatory} & \textbf{w} & \textbf{$\tau_{\rm min}$} & \textbf{w$(\rm op=50\%)$} & \textbf{w$(\rm op=100\%)$} & \textbf{D$_{\rm Major Axis}$} & \textbf{Comments}\\
   \textbf{(Telesc., Country)} & \textbf{(km)} & & \textbf{(km)} & \textbf{(km)}\\			    
\hline
\hline
Skalnate Pleso & 84.8 & 0.07 & 10.7 & 5.3 & +949 & > Roche limit\\
(1.30 m, Slovakia) & & & &  & & \\
\hline
Konkoly  & 2.1 & 0.60 & 1.9 & 1.0 & +843 & > Roche limit\\
(1.00 m, Hungary) & & & &  & & \\
\hline
Valle D'Aosta & 10.6 & 0.60 & 9.5 & 4.8 & +645 & > Roche limit\\ 
(0.81 m, Italy) & & & &  & & \\
\hline
Baronnies Proven\c{c}ales & 8.5 & 0.20 & 3.1 & 1.5 & +506 & $\sim$ Roche limit\\ 
(0.82 m, France)  & & & &  & & \\
\hline
Puimichel & 5.1 & 0.16 & 1.5 & 0.8 & +468 & < Roche limit\\ 
(1.04 m, France)  & & & &  & & \\
\hline
Calar Alto  & 16.5 & 0.03 & 1.1 & 0.5 & -307 & < Roche limit\\ 
(1.23 m, Spain)  & & & &  & & \\
\hline
Sierra Nevada  & 42.4 & 0.06 & 4.8 & 2.4 & -333 & < Roche limit\\
(1.50 m, Spain)  & & & &  & & \\
		\hline
	\end{tabular}}
\end{table*}

The other 22 negative datasets obtained did not provide enough constraints to confirm or rule out the presence of debris at different distances to 2002 GZ$_{32}$. We have constrained the widths of putative rings for different opacities using all the telescopes larger than 0.8 m outside 2002 GZ${32}$'s limb (see Table \ref{ringconstraints} and also Figure \ref{ring_geometries}). It is important to note that the existence of rings at distances larger than the Roche limit ($\sim$ 500-560 km for an object of 2002 GZ$_{32}$'s size with $\rho$ = 700-1000 kg/m$^3$, respectively) is very unlikely. This means that data from observatories at distances from the shadow-path centerline larger than this limit are not expected to provide constraints on potential rings. We can conclude that no rings or debris material of the kind found around Chariklo have been detected around 2002 GZ$_{32}$ by means of this stellar occultation, but narrow and thin rings with different geometries cannot be totally discarded.


\section{Conclusions}
\label{conclu}

\begin{itemize}

\item On 2017 May 20$^{\rm th}$ a multi-chord stellar occultation of a $m_{\rm V}$ = 14.0 mag star by the large centaur 2002 GZ$_{32}$ was detected from six telescopes in five different observatories located in Spain and Greece. The success of this occultation was possible thanks to a wide international campaign involving 31 telescopes\footnote{Two of the telescopes did not provide data due to bad weather conditions --Nunki Observatory in Greece-- and to technical problems --Kryoneri Observatory in Greece-- (Table \ref{SummaryObserv}).} located all over Europe (see Table \ref{SummaryObserv}). This is consistent with some of the empirical lessons learnt from $\sim$ 100 occultations by TNOs/Centaurs detected so far in the sense that a minimum of 15 observatories are typically needed in order to record a multi-chord occultation \citep{Ortiz2020}. 

\item A very elongated limb with axis (305 $\pm$ 17 km $\times$ 146 $\pm$ 8 km) was obtained from the best elliptical fit to the occultation data after applying time shifts. From this limb a geometric albedo of $p_{\rm V}$ = 0.043 $\pm$ 0.007 was derived. This albedo is higher than the radiometric albedos found using Herschel/Spitzer/ALMA thermal data, but compatible with them within error bars. From a rotational light curve obtained using data acquired close to the occultation date we determined that the occultation took place when 2002 GZ$_{32}$ was close to its minimum projected area, then the projected area obtained from the occultation directly provides the minimum area-equivalent diameter for this centaur.

\item Using the apparent axial ratio from the occultation (a'/b' = 2.09 $\pm$ 0.24), together with the rotational light curve constraints (i.e., the peak-to-peak amplitude, $\Delta$m = 0.13 $\pm$ 0.01 mag, and the rotational phase at the moment of the occultation), it is possible to derive a 3-D shape for 2002 GZ$_{32}$, obtaining an ellipsoid with axes, 366 km $\times$ 306 km $\times$ 120 km, as the best solution in terms of a $\chi^2$ minimization. It is also possible to constrain the aspect angle between 71.5$^\circ$ and 90.0$^\circ$ (minimum $\chi^2$ for an angle of 76$^\circ$). The 3-D model obtained is not consistent with a homogeneous body in hydrostatic equilibrium (assuming a Jacobi ellipsoid) for the known rotation period of 2002 GZ$_{32}$. Possible explanations for this non-hydrostatic equilibrium shape should be explored.

\item A mean area-equivalent diameter is derived from the maximum and minimum area-equivalent diameters obtained from the occultation ($D_{\rm eq_{max}}$ = 225 $\pm$ 12 km, $D_{\rm eq_{min}}$ = 211 $\pm$ 12 km) obtaining the value, <$D_{\rm eq}$> = 218 $\pm$ 12 km. This diameter is smaller than the equivalent diameter obtained from thermal models using observations acquired with Herschel/Spitzer/ALMA $D = 237^{+12}_{-11}$ km \citep{Lellouch2017}, but it is still compatible within error bars.

\item From the occultation light curves no dense ring or debris material comparable to the structures seen near Chariklo and Chiron were detected orbiting 2002 GZ$_{32}$. However, very narrow and/or optically thin rings around this centaur cannot be totally discarded.

\end{itemize}

\section*{Acknowledgements}
%
P.S-S. acknowledges financial support by the Spanish grant AYA-RTI2018-098657-J-I00 ``LEO-SBNAF'' (MCIU/AEI/FEDER, UE). P.S-S., J.L.O., N.M. and R.D. acknowledge financial support from the State Agency for Research of the Spanish MCIU through the ``Center of Excellence Severo Ochoa'' award for the Instituto de Astrof\'isica de Andaluc\'ia (SEV-2017-0709), they also acknowledge the financial support by the Spanish grant AYA-2017-84637-R and the Proyecto de Excelencia de la Junta de Andaluc\'ia J.A. 2012-FQM1776. The research leading to these results has received funding from the European Union's Horizon 2020 Research and Innovation Programme, under Grant Agreement no. 687378, as part of the project ``Small Bodies Near and Far'' (SBNAF). Part of the research leading to these results has received funding from the European Research Council under the European Community's H2020 (2014-2020/ERC Grant Agreement no. 669416 ``LUCKY STAR''). E.F-V. acknowledges funding through the Preeminant Postdoctoral Program of the University of Central Florida. Part of the data were collected during the photometric monitoring observations with the robotic and remotely controlled observatory at the University of Athens Observatory - UOAO \citep{Gazeas2016}. F.J.B. acknowledges financial support by the Spanish grant AYA2016-81065-C2-2-P. A.A-C. acknowledges support from FAPERJ (grant E26/203.186/2016) and CNPq (grants 304971/2016-2 and 401669/2016-5). A.C. acknowledges the use of the main telescope of the Astronomical Observatory of the Autonomous Region of the Aosta Valley (OAVdA). C.K. has been supported by the grants K-125015 and GINOP-2.3.2-15-2016-00003 of the National Research, Development and Innovation Office, Hungary (NKFIH). T.P. and R.K. acknowledge support from the project ITMS No. 26220120029, based on the Research and development program financed from the European Regional Development Fund and from the Slovak Research and Development Agency -- the contract No. APVV-15-0458. We are grateful to the CAHA and OSN staffs. This research is partially based on observations collected at Centro Astron\'omico Hispano-Alem\'an (CAHA) at Calar Alto, operated jointly by Junta de Andaluc\'ia and Consejo Superior de Investigaciones Cient\'ificas (IAA-CSIC). This research was also partially based on observation carried out at the Observatorio de Sierra Nevada (OSN) operated by Instituto de Astrof\'isica de Andaluc\'ia (CSIC). This article is also based on observations made with the Liverpool Telescope operated on the island of La Palma by the Instituto de Astrof\'isica de Canarias in the Spanish Roque de los Muchachos Observatory. This work has made use of data from the European Space Agency (ESA) mission {\it Gaia} (\url{https://www.cosmos.esa.int/gaia}), processed by the {\it Gaia} Data Processing and Analysis Consortium (DPAC, \url{https://www.cosmos.esa.int/web/gaia/dpac/consortium}). Funding for the DPAC has been provided by national institutions, in particular the institutions participating in the {\it Gaia} Multilateral Agreement. 

\section*{Data availability}

The data underlying this article will be shared on reasonable request to the corresponding author.




\bibliographystyle{mnras}

\bibliography{2002GZ32_occultation} 


\appendix
\section{Summary table including all the observatories and observers that supported the occultation campaign}
%

\begin{table*}
	\centering
	\caption{Observatories and observers that supported the stellar occultation by 2002 GZ$_{32}$. Sites ordered from north to south sorted by projected distance on the sky plane to the reconstructed centerline. Positive distances indicate sites north of the center line, negative distances indicate sites south of the center line.}
	\label{SummaryObserv}
	\begin{tabular}{cccccc} 
		\hline
\textbf{Observatory} 	& \textbf{Longitude (E)} & \textbf{Telescope}  & \textbf{Observer(s)} & \textbf{Observation}
& \textbf{Distance to}\\
\textbf{(Country)	}	& \textbf{Latitude (N)}  & \textbf{aperture} &    & 	& \textbf{centerline}				 \\
	& \textbf{Altitude (m)}  & 					   &  	   & & \textbf{(km)}\\
 \hline
 \hline
Borowiec Observatory  		& 17$^\circ$ 04' 28.6'' & 0.40 m  &  J. Horbowicz  &	Negative & +730.9\\
(Poland)				& 52$^\circ$ 16' 37.2''  &  & A. Marciniak & &\\   							    					
								& 123					  & 							& 		     & 			&\\
\hline
Smithy Observatory	& -00$^\circ$ 48' 58.2''  & 0.30 m  &  T. Haymes   &	Negative & +656.7\\
(UK)				& 51$^\circ$ 30' 23.8''  &  &  & & \\   							    					
								& 75					  & 							& 		     & 			&\\								
 \hline
Hamsey Green Observatory  		& -00$^\circ$ 04' 01.4''  & 0.28 m  &  M. Jennings   &	Negative & +651.5\\
(UK)				& 51$^\circ$ 19' 09.4''  &  &  & \\   							    					
								& 170					  & 							& 		     & 			&\\
								
\hline
PDlink Observatory 		& 18$^\circ$ 42' 09.5''  & 0.40 m  & P. Delincak &	Negative & +595.7\\
(Slovakia)				& 49$^\circ$ 24' 15.2''  &  &  & & \\   							    					
								& 680					  & 							& 		     & 			&\\	
\hline
Skalnate Pleso Observatory  		& 20$^\circ$ 14' 02.1'' & 1.30 m  &  T. Pribulla   &	Negative & +584.4\\
(Slovakia)				& 49$^\circ$ 11' 21.8''  &  & R. Kom\v{z}\'{i}k & & \\   							    					
								& 1826					  & 							& 		     & 			&\\								
\hline
Neutraubling Observatory  & 12$^\circ$ 12' 57.3''  & 0.28 m  & B. Kattentidt    &	Negative & +574.1\\
(Germany)				& 48$^\circ$ 59' 23.1''  &  &  & & \\   							    					
								& 333					  & 							& 		     & 			& \\
\hline	
Konkoly Observatory  		& 19$^\circ$ 53' 41.7''  & 1.00 m  &  A. Pal   &	Negative & +519.3\\
(Hungary)				& 47$^\circ$ 55' 06.0''  &  & Cs. Kiss & \\   							    					
							& 944					  & 							& 		     & 			& \\														
\hline
Landehen Observatory		& -02$^\circ$ 32' 16.0'' & 0.25 m  &  C. Ratinaud    &	Negative & +496.1\\
(France)				& 48$^\circ$ 26' 02.1''  &  &  & \\   							    					
								& 87					  & 							& 		     & 			&\\								
\hline
Valle D'Aosta Observatory  		& 07$^\circ$ 28' 42.0''  & 0.81 m  &  A. Carbognani   &	Negative & +396.8\\
(Italy)				& 45$^\circ$ 47' 22.0''  &  &  & \\   							    					
								& 1675					  & 							& 		     & 			& \\							
\hline
Baronnies Proven\c{c}ales Observatory  & 05$^\circ$ 30' 54.5'' & 0.82 m  &  M. Bretton   &	Negative & +311.4\\
(France)				& 44$^\circ$ 24' 29.3''  &  &  & \\   							    					
								& 820					  & 							& 		     & 			&\\								
\hline 
Monte Agliale Observatory	& 10$^\circ$ 30' 53.8'' & 0.50 m  &  F. Ciabattari   &	Negative & +300.3\\
(Italy)				& 43$^\circ$ 59' 43.1''  &  &  & \\   							    					
								& 758 					  & 							& 		     & 			&\\								
 \hline	 
Puimichel Observatory  		& 06$^\circ$ 01' 15.6'' & 1.04 m  &  J. Lecacheux   &	Negative & +288.0\\
(France)				& 43$^\circ$ 58' 48.7''  &  & S. Moindrot & & \\   							    					
								& 725					  & 							& 		     & 			& \\
\hline
T\'{e}lescope L\'{e}onard de Vinci	& 07$^\circ$ 04' 18.4'' &  0.40 m  &  F. Signoret   &	Negative & +268.6\\
(France)				& 43$^\circ$ 36' 15.7''  &  & A. Fuambu & & \\   							    					
								& 130					  & 			&				& 		     & 			\\								
 \hline
Latrape Observatory  		& 01$^\circ$ 17' 24.7'' & 0.35 m  &  J. Sanchez   &	Negative & +224.6\\
(France)				& 43$^\circ$ 14' 38.7''  &  &  & & \\   							    					
								& 363					  & 			&				& 		     & 			\\
\hline
Latrape Observatory  		& 01$^\circ$ 17' 24.4'' & 0.30 m  &  M. Boutet   &	Negative & +224.6\\
(France)				& 43$^\circ$ 14' 38.4''  &  &  & & \\   							    					
								& 355					  & 			&				& 		     & 			\\
\hline
Sabadell Observatory  		& 02$^\circ$ 05' 24.6'' & 0.50 m  & C. Perello    &	Negative & +123.9\\
(Spain)				& 41$^\circ$ 33' 00.2''  &  & A. Selva  & & \\   							    					
								& 224					  & 			&				& 		     & 			\\								
\hline
	\end{tabular}
\end{table*}

\begin{table*}
	\centering
	\contcaption{ - Summary table including all the observatories and observers that supported the occultation campaign}
\label{tab:continued.}
	\begin{tabular}{cccccc} 
		\hline
\textbf{Observatory} 	& \textbf{Longitude (E)} & \textbf{Telescope}  & \textbf{Observer(s)} & \textbf{Observation}
& \textbf{Distance to}\\
\textbf{(Country)	}	& \textbf{Latitude (N)}  & \textbf{aperture} &    & 	& \textbf{centerline}				 \\
	& \textbf{Altitude (m)}  & 					   &  	   & & \textbf{(km)}\\
\hline
\hline
Forcarei Observatory  		& -08$^\circ$ 22' 15.2''  & 0.50 m  &  H. Gonz\'{a}lez   &	Negative & +123.6\\
(Spain)				& 42$^\circ$ 36' 38.3''  &  & R. Iglesias-Marzoa & & \\   							    					
								& 670					  & 			&				& 		     & 			\\
\hline
Sant Esteve Observatory  		& 01$^\circ$ 52' 21.1''  & 0.40 m		& C. Schnabel  & Negative & +119.3\\
(Spain)				& 41$^\circ$ 29' 37.5''  & &    & & \\   							    					
								& 180					 & 			&				& 		  	 & 			     \\									
\hline 
Allariz Observatory & -07$^\circ$ 46' 13.0''	  & 0.25 m 	& L. P\'erez & Positive  & +102.5\\
(Spain)    & 42$^\circ$ 11' 57.0''    & 				& & & \\
				    & 514		  & 			&			& 		  & 		 \\ 
\hline
Javalambre Observatory & -01$^\circ$ 00' 58.6''	  & 0.40 m 				& R. Iglesias-Marzoa		 & Positive & +11.8\\
(Spain)        & 40$^\circ$ 02' 30.6''   & & J.L. Lamadrid    & & \\
				       & 1957           		  &					 & N. Ma\'{i}cas    & 			& \\

\hline 
Aras de los Olmos Observatory   & -01$^\circ$ 06' 05.4''	 & 0.52 m		&  	V. Peris  & 	Positive & +4.4\\
(Spain)        		& 39$^\circ$ 56' 42.0''     & &       & & \\
				       			& 1280           		     & 			&				& 		  	 &      \\
\hline
Nunki Observatory  		& 23$^\circ$ 30' 28.6'' & 0.40 m  &  N. Paschalis   &	Bad & -6.4\\
(Greece)				& 39$^\circ$ 10' 49.0''  &  &  & Weather & \\   							    					
								& 26					  & 			&				& 		     & 			\\   
 \hline
La Hita Observatory    & -03$^\circ$ 11' 09.8''  & 0.77 m	& N. Morales		 & Positive & -34.1\\
(Spain)        & 39$^\circ$ 34' 07.0''     &  & F. Organero      &  & \\
				       & 674              & 				&					 & 			     & 			\\
\hline				        
 
La Hita Observatory    & -03$^\circ$ 11' 09.8''  & 0.40 m 		&  L. Ana-Hernandez   & 	Positive, but with & -34.1\\
(Spain)        & 39$^\circ$ 34' 07.0''    & & F. Organero         & technical & \\
				       & 674           		  & 									& F. Fonseca & 	problems	 & \\
\hline 
Kryoneri Observatory  & 22$^\circ$ 37' 07.0'' & 1.20 m  &  J. Alikakos   &	Technical & -84.5\\
(Greece)				& 37$^\circ$ 58' 19.0''  &  &  & problems & \\   							    					
								& 930					  & 		&						& 		     & 			\\
\hline 
 
Univ. of Athens Observatory   & 23$^\circ$ 47' 00.1'' & 0.40 m	& 	K. Gazeas		 & Positive & -88.1\\
(Greece)        	  & 37$^\circ$ 58' 06.8''    &  & & & \\
				       		  & 250           		     & 			&				& 			     & 			\\												       
\hline				       
La Sagra Observatory  & -02$^\circ$ 33' 55.0''   & 0.36 m		   &  N. Morales &	Negative & -136.7\\
(Spain)				   & 37$^\circ$ 58' 56.6''   &    &  &  & \\   							    					
								& 1530					 & 			&				   & 			     & 			\\
\hline
La Murta Observatory  		& -01$^\circ$ 12' 10.0'' & 0.40 m  &  J.A. de los Reyes   &	Negative & -138.7\\
(Spain)						& 37$^\circ$ 50' 24.0''  &  & S. Pastor & & \\   							    					
								& 404					  & 			&				& 		     & 			\\ 		  					    
\hline
Adiyaman University Observatory  		& 38$^\circ$ 13' 31.5'' & 0.60 m  &  E. Sonbas   &	Negative & -182.0\\
(Turkey)				& 37$^\circ$ 45' 06.1''  &  &  & & \\   							    					
								& 700					  & 			&				& 		     & 			\\								
 \hline
Calar Alto Observatory  		& -02$^\circ$ 32' 54.1''  & 1.23 m  &  S. Mottola   &	Negative & -188.8\\
(Spain)				& 37$^\circ$ 13' 23.1''  &  & S. Hellmich & & \\   							    					
								& 2168					  & 			&				& 		     & 			\\
 \hline
Sierra Nevada Observatory & -03$^\circ$ 23' 05.0''  & 1.50 m  &  P. Santos-Sanz   &	Negative & -205.1\\
(Spain)				& 37$^\circ$ 03' 51.0''  &  & J.L. Ortiz & & \\   							    					
								& 	2896 				  & 							& 	N. Morales	     & 			& \\						
\hline
	\end{tabular}
\end{table*}


\bsp	
\label{lastpage}
\end{document}